\documentclass[sn-basic]{sn-jnl}% Basic Springer Nature Reference Style/Chemistry Reference Style
\usepackage{xcolor}
\usepackage{graphicx}%
\usepackage{multirow}%
\usepackage{amsmath,amssymb,amsfonts}%
\usepackage{amsthm}%
\usepackage{mathrsfs}%
\usepackage[title]{appendix}%
\usepackage{xcolor}%
\usepackage{textcomp}%
\usepackage{manyfoot}%
\usepackage{booktabs}%
\usepackage{algorithm}%
\usepackage{algorithmicx}%
\usepackage{algpseudocode}%
\usepackage{listings}%

\usepackage{graphicx}
\usepackage{caption}
\usepackage{subcaption}
\usepackage{geometry}
\usepackage{enumitem}
\theoremstyle{thmstyleone}%
\theoremstyle{thmstyletwo}%

\theoremstyle{thmstylethree}%

\begin{document}

\title[Article Title]{\large{\textbf{Scalable Field-Aligned Reparameterization for Trimmed NURBS}}}


\author[1]{\fnm{Zheng} \sur{Wei}}

\author*[1]{\fnm{Xiaodong} \sur{Wei}}\email{xiaodong.wei@sjtu.edu.cn}

\affil[1]{\orgdiv{The University of Michigan-Shanghai Jiao Tong University Joint Institute}, \orgname{Shanghai Jiao Tong University}, \orgaddress{ \city{Shanghai}, \country{China}}}


\abstract{In engineering design, one of the most daunting problems in the design-through-analysis workflow is to deal with trimmed NURBS (Non-Uniform Rational B-Splines), which often involve topological/geometric issues and lead to inevitable gaps and overlaps in the model. Given the dominance of the trimming technology in CAD systems, reconstructing such a model as a watertight representation is highly desired. While remarkable progress has been made in recent years, especially with the advancement of isogeometric analysis (IGA), there still lack a fully automatic and scalable tool to achieve this reconstruction goal. To address this issue, we present a semi-automatic and scalable reparameterization pipeline based on a scalable and feature-aligned meshing tool, QuadriFlow~\cite{ref:quadriflow}. On top of it, we provide support for open surfaces to deal with engineering shell structures, and perform sophisticated patch simplification to remove undesired tiny/slender patches. As a result, we obtain a watertight spline surface (multi-patch NURBS or unstructured splines) with a simple quadrilateral layout. Through several challenging models from industry applications, we demonstrate the efficacy and efficiency of the proposed pipeline as well as its integration with IGA. Our source code is publicly available on GitHub~\cite{ref:scaleuntrim}.
}

%%================================%%
%% Sample for structured abstract %%
%%================================%%




\maketitle
\section{Introduction}\label{sec1}

According to a study at Sandia National Laboratories in 2005~\cite{ref:sandia05}, the time spent to create analysis-suitable geometric models from CAD (Computer-Aided Design) models dominates the overall design-through-analysis process, which has become the de-facto bottleneck for the current software system to accommodate engineering designs with increasing scale and complexity~\cite{ref:hughes05}. The fundamental reason is that CAD systems ubiquitously adopt trimming for geometric modeling, which, however, is incompatible with the downstream applications such as CAE (Computer-Aided Engineering) and CAM (Computer-Aided Manufacturing). To address this interoperability challenge, researchers in both CAD and CAE have proposed to adopt a holistic view that encompasses the whole design-through-analysis process~\cite{ref:hughes05, ref:riesenfeld15}. In particular, isogeometric analysis (IGA) was proposed to fundamentally unify CAD and CAE by adopting the same CAD geometric models directly in CAE~\cite{ref:igabook}. It has gained enormous momentum from both academia and industry over the past decade.

Despite the remarkable advances IGA has made in various theoretical and methodological aspects, dealing with trimming remains an open and challenging problem. At its core, trimming is merely a mask scheme to hide part of a surface from users without changing the underlying mathematical description, leading to a geometric representation that does not conform to features like boundaries and creases. Moreover, representation of trimming curves is subject to tolerances specific to each individual CAD system. When transferring geometric data between different systems through a common exchange format (e.g., IGES, STEP), it may lead to loss of geometric accuracy, or even worse, topologically incorrect models~\cite{ref:marussig18}. As a consequence, CAD models are often visually ``intact" but fundamentally flawed with gaps and overlaps~\cite{ref:piegl05}, hence significantly hindering their direct adoption into the analysis procedure because analysis needs flawless and watertight geometric models.

Dealing with trimming is therefore of primary importance to achieve a seamless design-through-analysis process. The existing treatment can be divided into two categories: the geometry way and the analysis way. The geometry way seeks to reparameterize trimmed NURBS (Non-Uniform Rational B-Splines) while leaving the standard analysis procedure almost unchanged. In contrast, the analysis way appeals to novel boundary-unfitted methods that can perform analysis directly on trimmed CAD models. We focus on the geometry way in this work. Regarding the analysis way, interested readers may refer to a recent review~\cite{ref:deprenter23} and related key topics such as stabilization~\cite{ref:burman15, ref:puppi20, ref:wei21a}, numerical integration~\cite{ref:kudela16, ref:gunderman21, ref:garhuom22, ref:antolin22}, and boundary treatment~\cite{ref:ruess13, ref:wei21b}.

The geometry way either locally or globally reparameterizes trimmed NURBS. Local approaches only reparameterize regions around trimming curves to make them conform to geometric features while maintaining the rest of regions unchanged. For instance, T-splines were used to first convert every trimmed NURBS to an untrimmed T-spline surface and then stitch them together at the interfaces~\cite{ref:sederberg08}. Watertight Boolean operations were introduced to perform untrimming based on surface-to-surface intersections, but mesh refinement is needed to resolve such intersections and it needs to propagate through the entire model~\cite{ref:urick19}. The locality nature of such methods usually leads to a complex mesh structure and yields non-uniform distortion in the resulting parameterization throughout the model, i.e., high distortion near trimming curves and low distortion elsewhere.

In contrast, global approaches reparameterize a model entirely and can deliver high-quality models, which, however, inevitably involves quad\footnote{Abbreviated for quadrilateral.} meshing as an intermediate yet critical step. The primary goal of quad meshing is to capture a multi-block structure of the geometry such that each four-sided block can be filled with a regular quad mesh, where tensor-product-based splines (e.g., NURBS, Coons patches) have a natural fit~\cite{ref:hiemstra20, ref:shepherd22}. As a result, a trimmed model can be rebuilt into a multi-patch watertight representation. While a simple and clean block structure is always desired and may be obtained through, for instance, field-guided quad meshing~\cite{ref:bommes09, ref:bommes13, ref:campen15}, the process of finding it is usually time-consuming and difficult to scale because it involves solving a global optimization problem. A much less structured but easily scalable way is to convert triangular meshes into quad meshes by locally altering mesh connectivity~\cite{ref:tarini10, ref:remacle12}. Thanks to the recent advancement of unstructured spline technologies~\cite{ref:wei22}, even such meshes can serve as control meshes and yet yield globally smooth spline surface models. Nonetheless, quad meshing with a multi-block structure is particularly promising in handling trimming as it can greatly simplify the geometric representation without sacrificing geometric accuracy. However, with the ever-increasing scale and complexity of engineering designs, the scalability issue of direct quad meshing needs to be taken into account. Fortunately, the Instant Field-Aligned Meshes~\cite{ref:jakob15} and QuadriFlow~\cite{ref:quadriflow} have been proposed exactly for this purpose, where the global optimization problem is converted and handled by a Laplace-smoothing-like local iterative method.

In this work, we propose a semi-automatic and scalable pipeline to rebuild trimmed NURBS into a watertight spline model (e.g., multi-patch NURBS or unstructured spline model), with the source codes openly available on GitHub~\cite{ref:scaleuntrim}. The overall pipeline is shown in Fig.~\ref{fig:pipeline}. The input is a trimmed CAD model, which is assumed to be a 2-manifold with boundaries and may have topological/geometrical flaws (e.g., gaps). First, the CAD model is triangulated using an open-source tool such as OpenCASCADE~\cite{ref:occ} and Gmsh~\cite{ref:gmsh}. The resulting triangle mesh will have gaps or overlaps if the CAD model itself is problematic. A mesh repair step is followed to fix these issues and make it watertight. Next, based the repaired triangle mesh, we adopt QuadriFlow to rapidly generate a well-structured quad mesh, where we add new features in QuadriFlow to support open surfaces for mechanical models. Instead of directly serving as the control mesh, the quad mesh in this work is used to find the multi-block structure, or called a \emph{quad layout}. Then a patch extraction step is followed, where the entire quad mesh is divided into multiple patches and each patch is a regular and four-sided quad mesh. Usually, there might exist slender or tiny patches right after extraction. A patch simplification step is proposed to get rid of these redundant patches. A simple multi-block structure is then achieved, based on which we fit a NURBS surface to each quad patch. As quad patches conform at their interfaces, it is straightforward to obtain a watertight spline representation. To this end, the original trimmed CAD model is reparameterized entirely as a watertight spline model.

The contribution of the work is threefold:
\begin{enumerate}[topsep = 0.5 pt]
\item A semi-automatic, scalable, and modularized pipeline is proposed to convert a possibly ``leaky" trimmed CAD model to a watertight representation (e.g., multi-patch NURBS or unstructured splines).
\item QuadriFlow-based quad meshing is extended to support open surfaces with tailored treatment around boundaries.
\item A patch simplification method is proposed to remove the tiny or slender patches that result from quad meshing, with sharp features taken into account.
\end{enumerate}

The paper is organized as follows. 
Section~\ref{previous} discusses the pipeline design and reviews the closely related work. 
In Section~\ref{quad}, we introduce how to impose boundary constraints in QuadriFlow to support open surfaces.
Section~\ref{simplify} introduces how to simplify quad meshes to achieve a clean quad layout.
Several models from real-world problems are presented in Section~\ref{result} to demonstrate the efficacy and efficiency of the proposed method.
We conclude the work in Section~\ref{con}.
Two minor topics in terms of the contrition to the work, mesh repair and spline fitting, are covered in Appendices~\ref{app:fixing} and~\ref{app:fitting}, respectively.

\section{Pipeline design and previous work}\label{previous}
\begin{figure}[h]%
\centering
\includegraphics[width=1\textwidth]{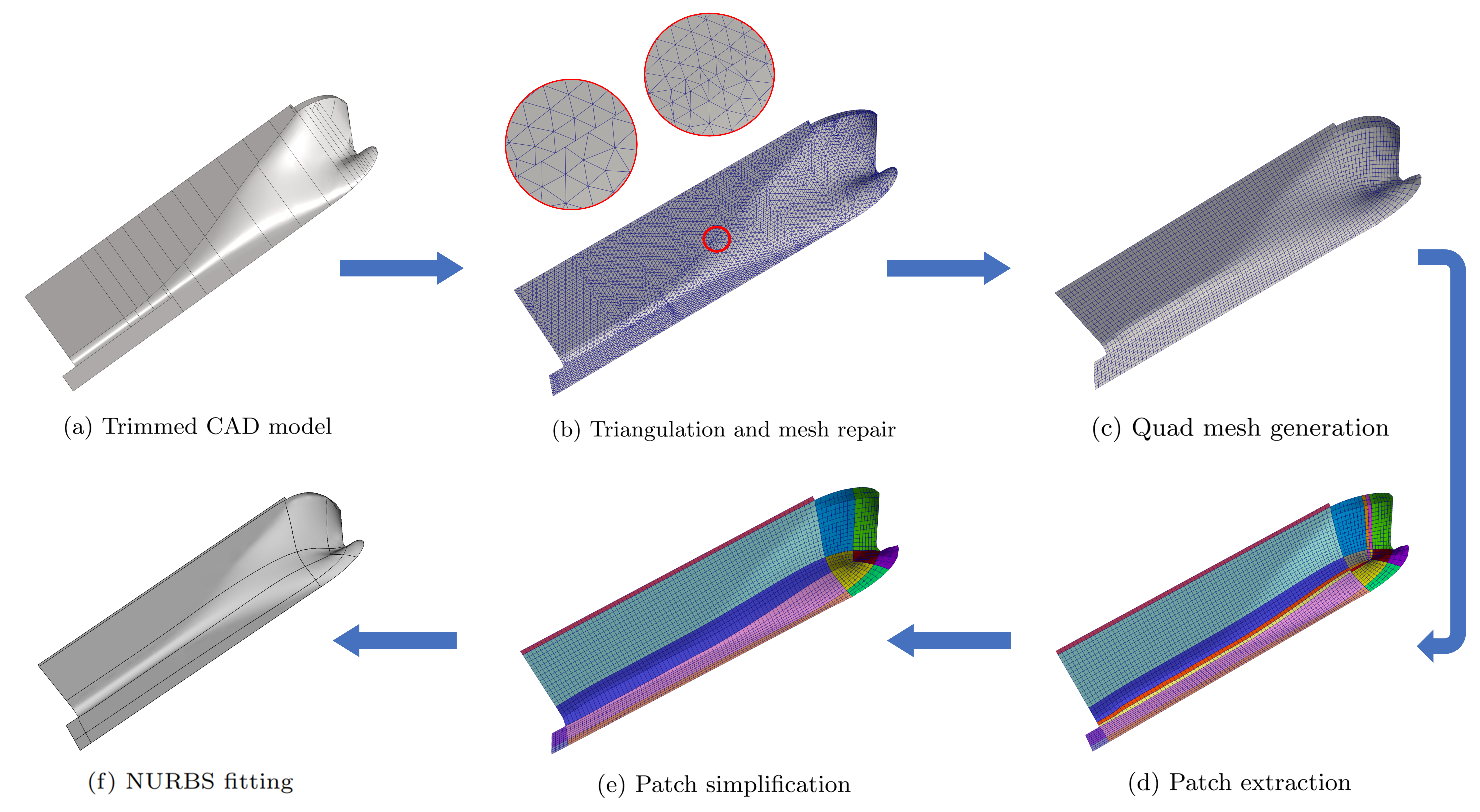}
\caption{The proposed pipeline framework to reconstruct a given trimmed CAD model entirely as a watertight representation.}
\label{fig:pipeline}
\end{figure}

The proposed pipeline framework is designed to be modularized, semi-automatic, and scalable. It is aimed to reconstruct a possibly ``leaky" trimmed CAD model as a watertight spline representation (e.g., multi-patch NURBS, unstructured splines), as shown in Fig.~\ref{fig:pipeline}. The reconstruction is a global operation applied to the entire model. The geometric error between the reconstructed model and the original one can be controlled through a user-defined tolerance. 

Given a trimmed CAD model, we first generate a corresponding triangle mesh using open-source software (e.g., OpenCASCADE, Gmsh). The resolution can be controlled by a certain tolerance. The triangle mesh serves an auxiliary purpose for the subsequent quad meshing. The input CAD model may be topologically/geometrically problematic due to trimming, leading to gaps and overlaps also in the triangle mesh. A mesh repair step is then followed to fix such defects, but it involves a huge variety of cases in practice. Among them, we only consider three common cases: duplicated vertices, a vertex lying on an edge or a face, and large gaps; see Appendix~\ref{app:fixing} for details.

With the watertight triangle mesh at hand, we proceed with the key step of the whole pipeline, quad meshing, which determines the structure and the quality of the reconstruction. For this task, we rely on an open-source tool, QuadriFlow, because it can produce well-structured meshes and more importantly, it is scalable, thus capable of accommodating large-scale inputs. However, QuadriFlow only supports closed surfaces. To work with shell structures in real-world engineering applications, we add new features to QuadriFlow to support open surfaces. 

With the quad mesh obtained from QuadriFlow, a set of four-sided patches can be readily extracted for the purpose of spline fitting, each of which is merely a regular grid. This set of patches gives a quad layout, which is desired to be simple in the sense that the number of patches is as small as possible. The quad layout is determined by the placement of singularities (i.e., extraordinary points where other than four edges meet). While being able to keep the number of singularities small, QuadriFlow does not guarantee the optimal placement of singularities, leading to possible clustering or misalignment of singularities and thus redundant patches. Therefore, a step of patch simplification is followed to remove such patches by modifying the mesh connectivity. Finally, we fit a NURBS surface to each patch (see Appendix~\ref{app:fitting} for details) with a user-specified tolerance to control the accuracy. This is done patch-wise, yet the final multi-patch NURBS representation is conforming across every interface because its conformality follows that of the quad mesh. To this end, we reparameterize the trimmed CAD model and obtain a watertight representation.

In what follows, we review the related work on mesh repair, quad meshing, patch simplification, and spline fitting/representation, pipeline design.

\subsection{Mesh repair}
CAD models often have invisible ``flaws" (such as gaps and overlaps) due to trimming~\cite{ref:piegl05}, the universal operation that enables flexible modeling of complex geometries. As a result, even the most sophisticated triangulation algorithms may fail to deliver a watertight triangle mesh from a ``leaky" CAD model. Therefore, a mesh repair step is needed, which involves a rich set of heuristic algorithms such as stitching gaps and removing overlaps.

Methods for triangle mesh repair are primarily divided into global and local methods. Global methods are more robust, typically reconstructing the entire model by re-meshing to eliminate various geometric defects~\cite{ju2004robust, hu2020fast}. However, global methods are time-consuming when dealing with large-scale models. Moreover, they may result in the loss of fine geometric features after reconstruction. Local methods target specific defects with specialized techniques~\cite{attene2010lightweight, chu2019repairing}. While not as robust as global methods, local methods are more efficient and better at preserving small features of the original model. We refer to~\cite{attene2013polygon, xiao2022automatic} for a comprehensive review of the field.

\subsection{Quad meshing}
Quad meshing plays a central role in the proposed pipeline as it provides the base structure of the final watertight representation and thus greatly determines the geometric quality. Quad meshing has been an active research area for the past two decades, with a focus on generating \textit{semi-regular} quad meshes that feature a multi-block structure. With the advance of IGA and its urgent need for analysis-suitable CAD models, semi-regular quad meshing has gained even more momenta because it has a perfect match with spline surfaces (e.g., NURBS). While we only touch on the most related work in what follows, interested readers may to refer to~\cite{ref:bommes_review} for a thorough review of the area.

%quad only, structured, alignment control, automatic, anisotropy 

In the literature, the prevailing way of semi-regular quad meshing falls into the family of parameterization-based methods, while polycube maps~\cite{ref:tarini04, ref:wangh08} and Centroidal Voronoi Tessellation~\cite{ref:levy10} also provide valuable alternatives. Parameterization-based methods can be further divided into two groups: direct global parameterization and field-guided methods. Direct global parameterization strives to construct a mapping from an input 3D surface to a 2D planar domain, with which a quad mesh can be readily generated by lifting a Cartesian grid in the 2D domain back to 3D. Typical methods in this category include conformal parameterization~\cite{ref:floater97, ref:gu03} that preserves angles and thus maintains orthogonality, harmonic parameterization~\cite{ref:tong06, ref:huang08} that is as-conformal-as-possible, and recent advances based on the surface foliation theory~\cite{ref:lei17a, ref:lei17b}, the Abel-Jacobi condition~\cite{ref:lei20, ref:zheng21}, and the Ricci flow~\cite{ref:shepherd22}.

On the other hand, field-guided methods provide explicit control over the desired local properties of a resulting quad mesh such as the orientation and the size of quad elements. This is achieved by carefully designing a so-called \textit{cross-field} (or \textit{frame field}), which consists of an orientation field and a sizing field. A typical subclass is a 4-rotationally-symmetric field~\cite{ref:ray08, ref:lai10} that represents orthogonal crosses with a uniform size. Field-guided methods involve two steps: cross-field generation and quad mesh synthesis. Cross-field generation, particularly orientation-field generation, aims to find the smoothest field subject to boundary conditions and sharp features. Two different formulations exist: a nonlinear formulation based on periodic functions~\cite{ref:palacios07, ref:ray09} and a mixed-integer formulation~\cite{ref:ray08, ref:bommes09, ref:panozzo12}. The key of this step is to automatically place the singularities (points lack of smoothness) in geometrically meaningful regions because it has a great impact on the quad mesh quality. Once a cross field is ready, quad mesh can be extracted either by explicitly tracing curves that align with the orientation field~\cite{ref:alliez03} or through global parameterization that respects the guiding field~\cite{ref:kalberer07, ref:bommes09,liao2014structure}. Explicit tracing often leads only to quad-dominant meshes (i.e, possibly with a few triangular faces). Global parameterization generally may not be bijective and thus leads to fold-overs, where various heuristics~\cite{ref:ebke13} as well as dedicated constraints~\cite{ref:hiemstra20} have been introduced to deal with the issue.

Almost all the parameterization-based methods needs to solve a global problem that depends on the entire mesh, and thus it is time-consuming and difficult to scale. In contrast, there exist two field-guided methods, Instant Field-Aligned Meshes~\cite{ref:jakob15} and QuadriFlow~\cite{ref:quadriflow}, that are based on local operators, so they are easy to implement and scalable to deal with large-scale models. Indeed, handling large scales is also one of the driving reasons for the need of IGA~\cite{ref:hughes05}. Instead of finding a global and continuous parameterization, these two local methods make use of discontinuous fields, an orientation field that guide the directions of quad edges and a position field that computes the positions of quad vertices, whose jumps are resolved by dedicated local smoothing operators. Moreover, they can achieve parameter-free alignment with sharp features by encoding the normal information in the to-be-minimized smooth energy, thereby providing a perfect match for the mechanical models. Compared to the Instant Field-Aligned Meshes, QuadriFlow eliminates the singularities in the position field by introducing additional regularity and non-fold-over constraints, and thus yields better structured meshes. Our pipeline is built upon QuadriFlow and we add boundary constraints in both fields to support open surfaces.

%depending on whether an auxiliary or intermediate mesh is needed.
\subsection{Mesh simplification}
While the placement of singularities plays a central role in determining the overall multi-block structure of a quad mesh, automatic quad meshing algorithms (e.g., QuadriFlow) may yield sub-optimal distribution of singularities, thereby leading to undesired redundant mesh blocks that are not aesthetically pleasing. Mesh simplification can serve as a remedy to achieve a clean quad layout by removing redundant patches while preserving the original geometric features (e.g., creases in mechanical parts). The connectivity of quad meshes inherently possesses global constraints, making it difficult to simply removing individual elements without affecting the global structure. Therefore, mesh simplification methods operate on at least one layer of elements. They can be categorized as local and global methods. Local methods mainly coarsen a quad mesh by deleting vertices and elements in local areas~\cite{daniels2009localized, bozzo2010adaptive}. Such methods are flexible and adaptive, but they may not reduce the number of singularities significantly. The current mainstream algorithms are global methods. For example, the dual structure of a mesh can be modified to perform simplification~\cite{shepherd2007topologic, daniels2008quadrilateral}, which is straightforward to implement and efficient for meshes with a large number of singularities. However, the resulting mesh is quad-dominated, thus requiring an additional subdivision step to obtain an all-quad mesh. Other viable global approaches start with an existing quad mesh and then alter the valence and position of singularities through various operations, such as collapsing and interpolating along specified paths~\cite{feng2021q, akram2022structure} and introducing separatrix constraints to connect the mis-aligned singularities~\cite{tarini2011simple, viertel2019coarse}. However, these methods need the input quad meshes to have high quality.

\subsection{Spline fitting and representation}
Once a quad mesh is ready, a corresponding watertight spline representation can be obtained through spline fitting~\cite{ref:nurbsbook, ref:floater03}. Different choices are available such as multi-patch NURBS and unstructured splines. In particular, unstructured splines hold the promise to accommodate the disparate needs for both design and analysis, thereby providing an ideal candidate for IGA. Unstructured splines consist of a large family of methods and can be broadly divided into two groups based on whether spline functions find a finite representation or not. In the group of finite representation, multi-patch $G^1$ splines~\cite{ref:collin16, ref:kapl18} and unstructured $C^1$ splines based on degenerated B\'ezier patches~\cite{ref:toshniwal17, ref:casquero20, ref:wei22} are two typical examples and have gained considerable attentions in recent years. The finite representation is particularly beneficial when integrating CAD with CAE through B\'ezier extraction~\cite{ref:borden11}. On the other hand, the group of infinite representation indicates subdivision surfaces~\cite{ref:burkhart10, ref:wei15, ref:wei16, ref:pan16}, which offer great flexibility and efficiency in modeling complex surfaces and have wide applications in computer animation and CAD. Some of the recent efforts along this direction are dedicated to addressing the issues with surface quality and the approximation property~\cite{ref:lix19, ref:wei21c}.

In summary, a semi-automatic pipeline that allows for a  seamless integration of above methods is highly desired to accommodate the needs of various geometric modeling and analysis scenarios.
Such pipeline frameworks have been studied for both surfaces and volumes. The one proposed in~\cite{lai2017integrating} is among the first of such pipelines, which is based on different variants of T-splines and further integrated with ABAQUS through B{\'e}zier extraction for industrial applications. A follow-up work based on polycube maps was dedicated to the reconstruction of watertight volumetric spline models and the integration with LS-DYNA~\cite{yu2022hexgen}. However, these two pipelines often involve user intervention during reconstruction and thus can be time-consuming in challenging problems. On the other hand, based on the frame-field guided parameterization, a semi-automatic pipeline was proposed to reconstruct a trimmed CAD model (assumed to be topologically/geometrically correct) as a set of Coon's patches~\cite{ref:hiemstra20}, where particular care is taken to prevent fold-over around singularities. Another pipeline was proposed based on a different way of parameterization through Ricci flow~\cite{shepherd2022feature}, where an optimization technique was extended to handle arbitrary path constraints and ensure valid parameterizations around singularities. While these methods feature high-quality and analysis-suitable reconstructions, the underlying parameterization methods need to repeatedly solve global problems, thereby hindering their scalability in dealing with large-scale models.

\section{QuadriFlow-based quad meshing for open surfaces}\label{quad}
After triangulation of the input CAD model and the subsequent mesh repair (see Appendix~\ref{app:fixing} for details), we now have a watertight triangle mesh at our disposal. Based on this, we adopt QuadriFlow~\cite{ref:quadriflow}, an open-source tool with a scalable algorithm, for quad remeshing to obtain a well-structured quad mesh, so that eventually we can convert it to a spline representation.

\begin{figure}[h]
\centering
\includegraphics[width=1\textwidth]{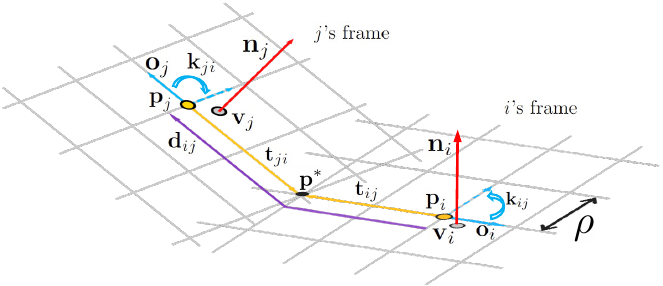}
\caption{Terminology illustration. \(\textbf{v}_i\) and \(\textbf{v}_j\) are two neighboring vertices in the triangle mesh, where gray lines indicate the uniform local grids in the tangent planes, and red arrows are unit normals \(\textbf{n}_i\) and \(\textbf{n}_j\). \(\textbf{o}_i\) and \(\textbf{o}_j\) are representative directions that can be matched in the same direction by rotation matrices \(\textbf{k}_{ij}\) and \(\textbf{k}_{ji}\). \(\textbf{p}_i\) and \(\textbf{p}_j\) are origins of local grids that can be made coincide on \(\textbf{p}^*\) by integer translations \(\textbf{t}_{ij}\) and \(\textbf{t}_{ji}\), respectively. The integer offsets \(\textbf{d}_{ij}\) is the ``distance'' from \(\textbf{p}_i\) to \(\textbf{p}_j\). \(\rho\) is the user-defined grid spacing.}\label{illustration}
\end{figure}

QuadriFlow is developed based on the seminal work of Instant Field-Aligned Meshes~\cite{ref:jakob15}. Instant Field-Aligned Meshes compute two fields for the parameterization purpose: an orientation field that guides the edge directions in the resulting quad mesh, and a position field that yields the vertex positions in the quad mesh. Compared to other field-aligned parameterization methods, Instant Field-Aligned Meshes feature a collection of discontinuous local parameterizations. Therefore, small local problems are solved instead of complex global problems, providing the possibility for developing scalable algorithms. Two kinds of singularities arise when either of the two fields is not smooth. While singularities of the orientation field are intrinsic to geometric models and cannot be removed in general, QuadriFlow succeeds in removing the singularities in the position field, which stand out as T-junctions and represent the transition of different mesh resolutions.

\subsection{Orientation field with boundary alignment}\label{subsec41}
We define an input triangle mesh \textit{M}= (\textit{V}, \textit{E}, \textit{F}), where  \( \textit{i} \in \textit{V} \) is a vertex index with position \( \textbf{v}_i \in \) $ \mathbb{R}^3 $, \( \textit{E} \subset \textit{V} \times \textit{V} \)  is the set of edges, and \textit{F} is the set of faces. The \textit{orientation field} is defined as a four-way rotationally symmetric field in tangent planes, which is computed first. It is composed of four mutually perpendicular unit vectors, which resemble a cross (therefore also referred to as a \textit{cross field}). As its name suggests, it can guide the edge directions in the resulting quad mesh. As the orientation field is rotationally symmetric, for every vertex \( \textit{i} \in \textit{V} \), we pick a \textit{representative direction} \( \textbf{o}_i \) in its tangent plane. This vertex-based method is beneficial when imposing feature constraints, where we can modify the orientation field to align with geometric features (see details later). The orientation field at vertex \textit{i} can be represented by rotating \( \textbf{o}_i \) counterclockwise by \(\pi/2\) three times around its normal \( \textbf{n}_i \). Therefore, the orientation field is denoted as \( \textbf{R}_3 \)(\( \textbf{n}_i, \textit{k} \))\( \textbf{o}_i \), where \( \textbf{R}_3 \)(\( \textbf{n}_i, \textit{k}\)) is the rotation matrix in the three dimensional space by \textit{k}\(\pi/2\) about \( \textbf{n}_i \), \( \textit{k} \in  \{0,1,2,3\} \). The \textit{smoothness energy} is introduced to measure the difference between two neighboring representative directions directly in the geometry space,
\begin{equation}\label{eq1}
E_O(\textbf{o},\textit{k})=\sum\limits_{i\in \textit{V}}\sum\limits_{j\in \textit{N}_i}\angle \ (\textbf{R}_3( \textbf{n}_i, \textit{k}_{ij}) \textbf{o}_i, \ \textbf{R}_3( \textbf{n}_j, \textit{k}_{ji})\textbf{o}_j)^{2},
\end{equation}
where \( \textit{N}_i \) is the set of adjacent vertices to vertex \textit{i}, \( \angle(\textbf{a},\textbf{b})\) is the angle between vectors \textbf{a} and \textbf{b}, and \( k_{ij} \in \{0,1,2,3\} \) indicates the number of \(\pi/2\)-rotations for \( \textbf{o}_i \) to align with \( \textbf{o}_j \) (likewise for \( k_{ij}\)). An optimal orientation field, i.e., when the energy defined in Eq.~\eqref{eq1} is minimized, aligns each pair of adjacent representative directions as closely as possible. To address the mixed-integer problem that arises from real variables \( \textbf{o}_i \), \( \textbf{o}_j \) and integer variables \( k_{ij}\),  \( k_{ji}\), Instant Field-Aligned Meshes propose so-called \textit{local Gauss-Seidel iteration}. Both kinds of variables are first computed as real numbers. Next, \( k_{ij}\) and \( k_{ji}\) are rounded to their nearest integers, which only affect certain real variables (e.g., \( \textbf{o}_i \) and \( \textbf{o}_j \)) in a local region. Such real variables are then updated by Gauss-Seidel iteration. The whole optimization process is done iteratively and locally, which has a similar form to the Laplace-smoothing operator~\cite{taubin1995signal}, 
\begin{equation}
\textbf{o}_i \gets \textbf{o}_i +\sum_{j\in \textit{N}_i}\omega_{ij}\ \textbf{R}_3( \textbf{n}_i, \textit{k}_{ij}) \textbf{o}_j, \ \ \ \  \ \ \ \textbf{o}_i \gets \textbf{o}_i / \Vert \textbf{o}_i\Vert,
\end{equation}
where $\omega_{ij}$ is a weight and \(\omega_{ij}\) = 1 is usually adopted for uniform meshes. In the end, the minimizers in terms of representative directions and integer rotations can be achieved,

\begin{equation}
\textbf{o}^*, k^* = \underset{\textbf{o}, k}{\arg\min}\ E_O(\textbf{o}, k).
\end{equation}

\begin{figure}[h]%
\centering
\includegraphics[width=1\textwidth]{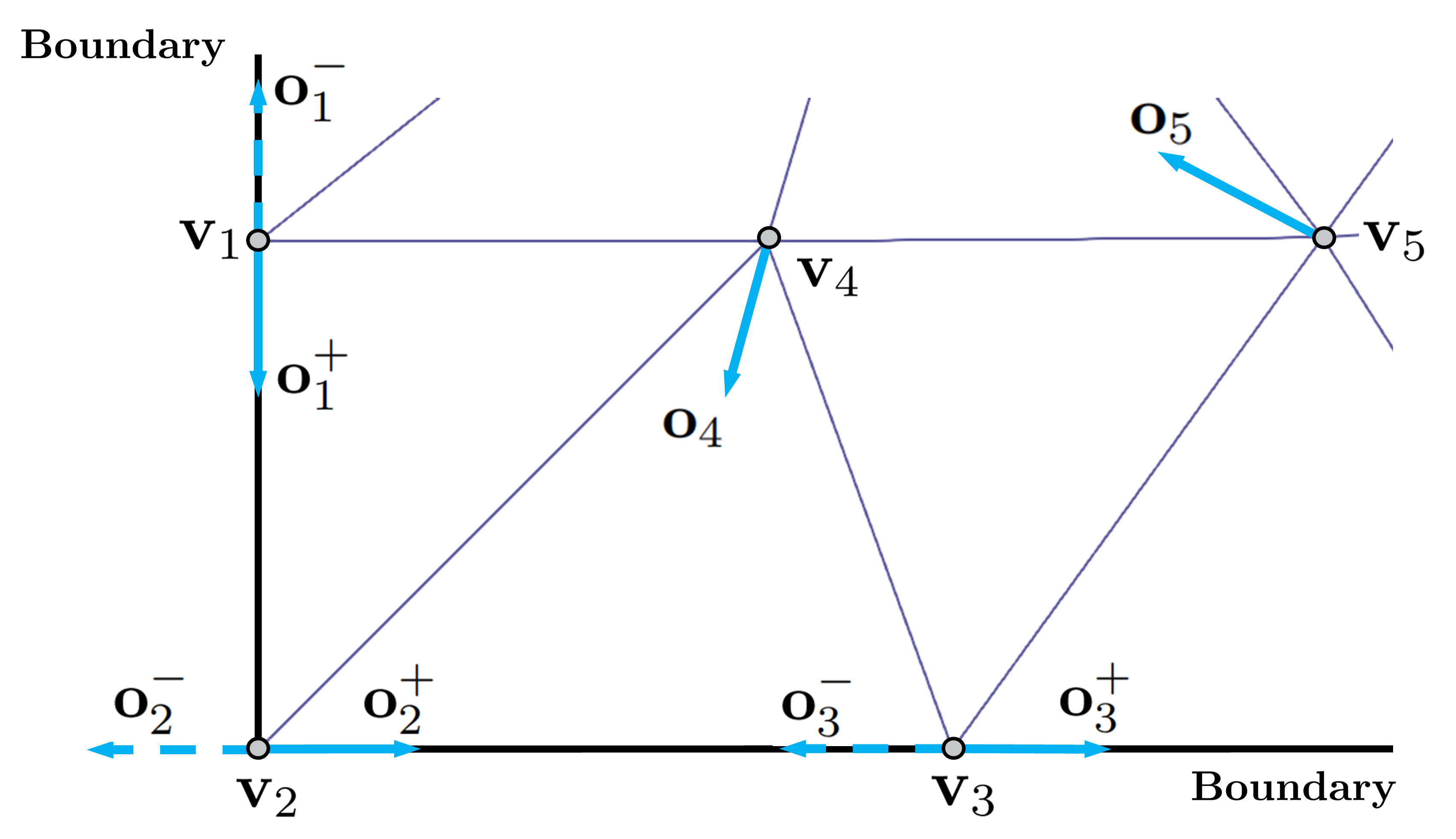}
\caption{The boundary constraint in the orientation field: the representative direction needs to align with the boundary tangent. When a boundary vertex is a corner (e.g., $\mathbf{v}_2$), its representative direction is prescribed along one of the two boundary edges (e.g., $\mathbf{o}_2^+$ or $\mathbf{o}_2^-$). For a non-corner boundary vertex (e.g., $\mathbf{v}_1$, $\mathbf{v}_3$), its representative direction is prescribed along the boundary tangent and it can only take two possible directions, e.g., \(\textbf{o}_{1}^+\) or \(\textbf{o}_{1}^-\) for $\mathbf{v}_1$. For interior vertices, their representative directions (e.g., \(\textbf{o}_{4}\), \(\textbf{o}_{5}\)) are free and to be determined by optimization.}
\label{ofb}
\end{figure}
  
On the other hand, we need to impose boundary constraints in the orientation field to support open surfaces. We first identify whether a boundary vertex is a corner or not depending on the geometric information. A boundary vertex is a corner if the internal angle between its two neighboring boundary edges is within 
 \((0,\pi-\theta]\) or \([\pi+\theta, 2\pi\)), where \(\theta \) is a given threshold (e.g., \(\pi/4\)). Otherwise, it is not a corner. 
For a corner, during optimization we fix its representative direction as the direction of one of its boundary edges. For a non-corner boundary vertex, throughout optimization its representative direction is prescribed as its tangent (or the opposite direction); see Fig.~\ref{ofb}. The tangent is computed by a weighted combination of the directions of its two boundary edges.

\begin{algorithm}
\caption{Minimization of the smoothness energy for the orientation field}\label{algo1}
\begin{algorithmic}[1]
\State Initialize the representative directions $\{\textbf{o}_i\}$ randomly;
\State Divide vertices \textit{V} into two groups: boundary vertices $\textit{V}_B$ and interior vertices$\textit{V}_I$;
\For{$\mathrm{iter} = 0$ \textbf{to} $\mathrm{max}\_\mathrm{iteration} $}
    
    \For{$\mathrm{every \ boundary \ vertex} \ i \in \textit{V}_B$}
    \State $\mathrm{Find \ its \ adjacent \ vertex }$ $\textbf{v}_j$ $\mathrm{that \ shares \ an \ edge \ with}$ $\textbf{v}_i$;
    \State $\textit{k}_{ij} = {\arg\min}\ \angle( \textbf{R}_3( \textbf{n}_i, \textit{k} ) \textbf{o}_i, \textbf{R}_3( \textbf{n}_j, \textit{k} )\textbf{o}_j)$;
    \State $\textbf{o}_i \gets \textbf{o}_i +\sum_{j\in \textit{N}_i}\omega_{ij}\textbf{R}_3( \textbf{n}_i, \textit{k}_{ij}) \textbf{o}_j$;
    \State $\textbf{o}_i \gets \textbf{o}_i / \Vert \textbf{o}_i\Vert$;
    \State Impose boundary constraints on $\textbf{o}_i$;
    \EndFor
    \For{$\mathrm{every \ interior \ vertex} \ i \in \textit{V}_I$}
    \State $\mathrm{Find \ its \ adjacent \ vertex }$ $\textbf{v}_j$ $\mathrm{that \ shares \ an \ edge \ with}$ $\textbf{v}_i$;
    \State $\textit{k}_{ij} = {\arg\min}\ \angle( \textbf{R}_3( \textbf{n}_i, \textit{k} ) \textbf{o}_i, \textbf{R}_3( \textbf{n}_j, \textit{k} )\textbf{o}_j)$;
    \State $\textbf{o}_i \gets \textbf{o}_i +\sum_{j\in \textit{N}_i}\omega_{ij}\textbf{R}_3( \textbf{n}_i, \textit{k}_{ij})\textbf{o}_j$;
    \State $\textbf{o}_i \gets \textbf{o}_i / \Vert \textbf{o}_i\Vert$;
    \EndFor
\EndFor
%\State push a boundary $\textbf{o}_i^*$ into queue
%\While{(not queue empty)}
%\State $\textbf{o}^* = $ pop(queue)
%\State match adjacent $\textbf{o}_j^*$ with $\textbf{o}^*$
%\State push unrepeated $\textbf{o}_j^*$ into queue
%\EndWhile

\end{algorithmic}
\end{algorithm}

\begin{figure}[h]
  \subcaptionbox{Inconsistent representative directions}[0.45\textwidth]{\includegraphics[width=1.2\linewidth]{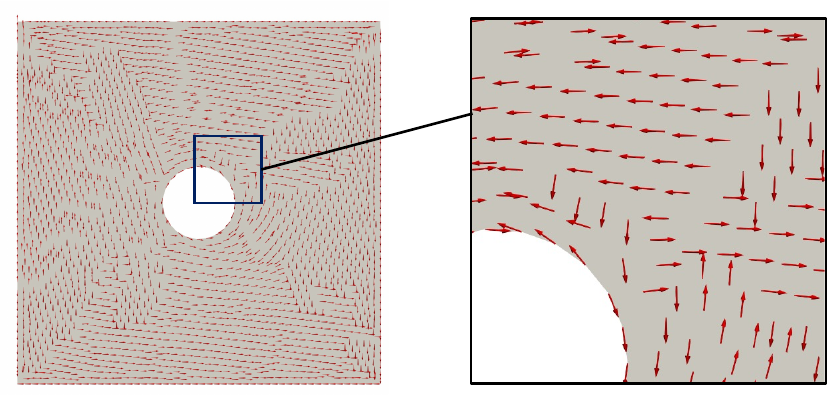}}
  \hfill
  \subcaptionbox{Inconsistent representative directions}[0.45\textwidth]{\includegraphics[width=0.55\linewidth]{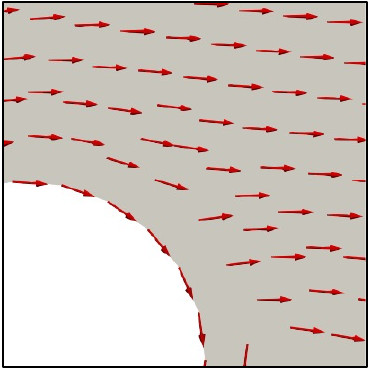}}
  \caption{Field matching of representative directions. (a) Inconsistent representative directions \(\textbf{o}_i^*\) before field matching, and (b) consistent representative directions after field matching.}
  \label{ofm}
\end{figure}

When optimization is done, we obtain a set of optimal representative directions \(\textbf{o}_i^*\). However, these directions generally do not match in the sense that two adjacent directions may differ more than an angle of \(\pi/4\); see Fig.~\ref{ofm}(a). Under such an inconsistent alignment of representative directions, there exits a high risk of mesh quality degradation or even mesh fold-over~\cite{vaxman2016directional}. To prevent such an issue, it is crucial to reinstate a consistent alignment of representative directions, which can be effectively resolved by field matching, as mentioned in ~\cite{ref:hiemstra20, viertel2019coarse}. An example after field matching is shown in Fig.~\ref{ofm}(b).

\subsection{Position field with boundary alignment}\label{subsec42}
Based on the result of the orientation field, we proceed to compute the position field on the triangle mesh \textit{M}. Now each vertex \textit{i} is associated with a cross (i.e., a set of four-way vectors). We pick two perpendicular directions, \( \textbf{o}^*_i \) and \( \textbf{R}_3 \)(\( \textbf{n}_i, 1\))\( \textbf{o}^*_i \), as the axes to set up a local frame in its tangent plane and a corresponding lattice with a uniform spacing of \(\rho\). More precisely, the local lattice is written as
%After computing orientation field \( \textbf{o}^* \), the local parameterization is generated for each vertex \textit{i} in integer tangent plane with the origin \( \textbf{p} \in \) $ \mathbb{R}^3 $ and basis directions \( \textbf{o}^*_i \) and \( \mathrm{rot}_3  \)(\( \textbf{n}_i, \frac{\pi}{2}\))\( \textbf{o}^*_i \). 

\begin{equation}
\Gamma(\textbf{p}_i,\textbf{n}_i,\textbf{o}^*_i,\textbf{t}_i ) = \textbf{p}_i+\rho\sum\limits_{k=0}^{1} \mathrm{t}_{i,k}\textbf{R}_3( \textbf{n}_i, \textit{k}) \textbf{o}^*_i,
\end{equation}
where \( \textbf{p}_i \in \) $ \mathbb{R}^3 $ is the origin of the local lattice, and \( \textbf{t}_i = (\mathrm{t}_{i,0}, \mathrm{t}_{i,1}) \in \mathbb{Z}^2 \) 
represents the integer translations in the two directions \( \textbf{o}^*_i \) and \( \textbf{R}_3  \)(\( \textbf{n}_i, 1\))\( \textbf{o}_i^* \). Note that the local lattice is invariant upon integer translations. The parameter \(\rho\) controls the density of the resulting quad mesh. It is mainly determined by the geometric features and can be tuned by users within a certain range. Every vertex \textit{i} is associated with a lattice origin \( \textbf{p}_i\), together yielding the position field, which is the target to be determined.

We aim to find an optimal position field in the sense that for every pair of adjacent vertices, their local lattices ``overlap" as much as possible. For every vertex \textit{i}, its local lattice is uniquely determined by the origin \( \textbf{p}_i\). The overall difference among local lattices is quantified as
\begin{equation} \label{eq5}
E_P(\textbf{p},\textbf{t})=\sum\limits_{i\in \textit{V}}\sum\limits_{j\in \textit{N}_i}\Vert \ \Gamma(\textbf{p}_i,\textbf{n}_i,\textbf{o}^*_i,\textbf{t}_{ij})- \Gamma(\textbf{p}_j,\textbf{n}_j,\textbf{o}^*_j,\textbf{t}_{ji} )\Vert^{2}_2,
\end{equation}
%where \( \textbf{t}_{ij} \) and \( \textbf{t}_{ji} \), removing transformation ambiguities, represent integer variables in the adjacent tangent plane.   
where \( \textbf{t}_{ij} \) (likewise for \( \textbf{t}_{ji} \)) indicates the integer translation of the local lattice of vertex \textit{i} to best match that of vertex j; see Fig.~\ref{illustration} for illustration. These integer translations \( \textbf{t}_{ij} \) and \( \textbf{t}_{ji} \) are introduced to guarantee a unique \( \textbf{p}_i\) with \(\Vert \textbf{p}_i - \textbf{v}_i \Vert < \rho\). Otherwise, there will be infinite many solutions for valid \( \textbf{p}_i\), e.g., \(\textbf{p}_i + \rho k\textbf{o}^*_i,  k\in \mathbb{Z} \). 
As this optimization problem involves real variables \( \textbf{p}_{i} \) and integer variables \(\textbf{t}_{ij}\), it is again a mixed-integer problem. Instant Field-Aligned Meshes adopt the same local Gauss-Seidel iteration as in the case of the orientation field to find minimizers. A similar form to Laplace smoothing is also used to resolve the position field in a local and iterative manner,
\begin{equation}
\textbf{p}_i \gets \textbf{p}_i +\sum\limits_{j\in \textit{N}_i}\omega_{ij}\ \Gamma(\textbf{p}_{ji},\textbf{n}_i,\textbf{o}^*_i,\textbf{t}_{ij}), \ \ \ \ \textbf{p}_i \gets \mathrm{round}(\textbf{p}_i / \substack{\sum_{j\in \textit{N}_i}} \omega_{ij}),
\end{equation}
where \( \textbf{p}_{ji} \) represents the description of \( \textbf{p}_{j} \) in \(\textbf{v}_i\)'s frame. Through this process, the minimizers can be obtained,
\begin{equation}
\textbf{p}^*, \textbf{t}^*=\underset{\textbf{p},\textbf{t}}{\arg\min}\ E_P(\textbf{p},\textbf{t}). 
\end{equation}
%\newline Since all these vertices can be merged to integer coordinates, the length of edges in the generated quad mesh will be close to the \(\rho\).

When the overall difference in Eq.~\eqref{eq5} is minimized, the local lattices will overlap the most. For example, in Fig.~\ref{illustration}, through \( \textbf{p}_i\), \( \textbf{t}_{ij} \) (associated with \( \textbf{v}_i\)) and \( \textbf{p}_j\), \( \textbf{t}_{ji} \) (associated with \( \textbf{v}_j\)), \( \textbf{v}_i\) and \( \textbf{v}_j\) will be found to correspond to the same vertex in the quad mesh; see the intersection point (marked in black). Computation of the position field can be seen as a ``clustering" procedure, where each cluster corresponds to a quad-mesh vertex. Moreover, every cluster can only locate at a grid point in a local lattice, thus having a pair of integer coordinates (i.e., multiples of \(\rho\)). As a result, the edge length of the quad mesh is uniform and close to \(\rho\).

To support open surfaces, we need to impose boundary constraints also in the position field. During optimization, lattice origins need to satisfy the following two conditions:
\begin{enumerate}[topsep = 0.5 pt]
\item For a boundary vertex \(\textbf{v}_i\) that is not a corner, its corresponding \(\textbf{p}_i\) must still lie on the boundary. For \(\textbf{v}_i\) that is a corner, its corresponding \(\textbf{p}_i\) must coincide with \(\textbf{v}_i\). 
\item For a interior vertex \(\textbf{v}_i\) that is close to the boundary, its corresponding \(\textbf{p}_i\) cannot move beyond the boundary.
\end{enumerate}
\vspace{5pt}

\begin{figure}[h]
  \subcaptionbox{Treatment of boundary vertices}[0.44\textwidth]{\includegraphics[width=1\linewidth]{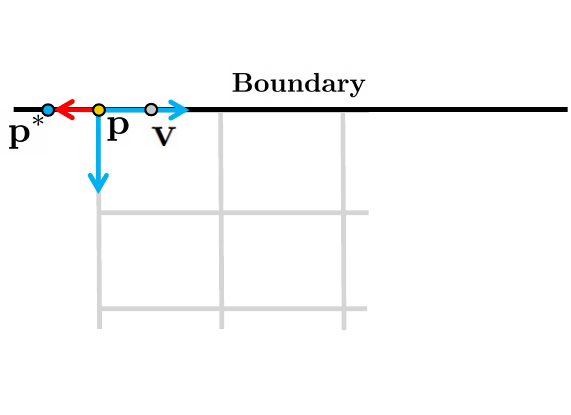}}
  \hfill
  \subcaptionbox{Treatment of interior vertices}[0.4\textwidth]{\includegraphics[width=1\linewidth]{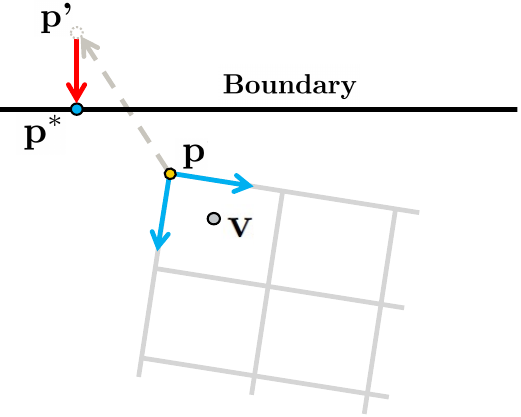}}
  \caption{Boundary constraints on the position field. (a) The lattice origin of a boundary vertex that is not a corner can only move along the boundary. (b) The lattice of an interior vertex near the boundary cannot go beyond the boundary; otherwise, the origin will be projected back onto the boundary.}
  \label{pfb}
\end{figure}

These two conditions are illustrated in Fig.~\ref{pfb}. To ensure Condition 1, the lattice origin of a boundary vertex is constrained to only move along the boundary. To satisfy Condition 2, when the lattice origin of an interior vertex goes beyond the boundary, it will be projected back onto the boundary. However, these constraints may introduce undesired singularities on the boundary, so next we discuss how to eliminate them.

\subsection{Singularity control near the boundary}\label{subsec43}
Singularities arise when either of the two fields lacks smoothness. While singularities in the orientation field are necessary for complex geometries, those in the position field merely indicate the transition of different mesh resolutions and can be removed~\cite{ref:quadriflow}. QuadriFlow introduced several dedicated constraints for this purpose. However, when we add new boundary constraints to the position field, a violation of these singularity-free conditions would occur in general, so we need a corresponding fix.

The singularities in the position field are identified through the so-called \emph{integer offset} \( \textbf{d}_{ij} \), which is the ``distance'' traveling from \( \textbf{p}_{i} \) to \(\textbf{p}_{j} \) when restricted onto the local lattices (see Fig.~\ref{illustration}). It is defined as

\begin{equation}
\textbf{d}_{ij}=  \textbf{t}_{ij} - \textbf{R}_2(k_{ij},k_{ji})\textbf{t}_{ji},
\end{equation}
where \(\textbf{R}_2\)(\(k_{ij}\),\(k_{ji}\)) is a two-dimensional rotation matrix with an angle of \((k_{ij}-k_{ji})\)\(\pi/2\) (counterclockwise), and it is introduced to transform the local frame of \( \textbf{v}_{j} \) to match that of \( \textbf{v}_{i} \), such that \( \textbf{t}_{ji} \) and \( \textbf{t}_{ij} \) are described in the same frame of \( \textbf{v}_{i} \). Position field singularity is detected when the sum of the integer offsets for a triangle is not zero,  
\begin{equation}
\textbf{d}_{ij} + R_{jk}^i\textbf{d}_{jk}+R_{ki}^i\textbf{d}_{ki} \neq \textbf{0} \ \ \ \ \ \ \forall \bigtriangleup_{ijk}  \in F,
\end{equation}
where \(R_{jk}^i = \textbf{R}_2(k_{jk},k_{kj})\) and \(R_{ki}^i = \textbf{R}_2(k_{ki},k_{ik})\) are rotation matrices in \(\textbf{v}_i\)'s frame.

QuadriFlow introduced two sets of constraints on the integer offsets: \textit{regularity constraints} that are aimed to remove the singularities in the position field, and \textit{consistency constraints} that ensure no-fold-over (i.e., positive Jacobian); see Eqs.~\eqref{eq10} and~\eqref{eq11} respectively. Solving such a mixed-integer constrained optimization problem is NP-hard. QuadriFlow addresses this challenging problem through a heuristic approach. It first finds the minimizers for \( \textbf{p}_i^* \) and \( \textbf{t}_{ij}^* \) without considering any constraints. Then it adjusts the corresponding \(\textbf{d}^*\) (computed using \(\textbf{p}_i^*\) and \(\textbf{t}_{ij}^*\)) with a minimum change to satisfy the two sets of constraints. More specifically, the following constrained optimization needs to solved,
\begin{align}
\label{eqmind}
\underset{\textbf{d}}{\min} \ \ \ &\|\mathbf{d} - \mathbf{d}^*\|_1 \\
\label{eq10}
\mathrm{subject} \ \mathrm{to} \ \ \ &\textbf{d}_{ij} + R_{jk}^i\textbf{d}_{jk}+R_{ki}^i\textbf{d}_{ki} = \textbf{0} \  \ \ \ \ \ \forall \bigtriangleup_{ijk}  \in F, \\
\label{eq11}
&\mathrm{det}[\textbf{d}_{ij},R_{ki}^i\textbf{d}_{ki}] \geq \textbf{0} \  \ \ \ \ \ \forall \bigtriangleup_{ijk}  \in F. 
\end{align}
Once the optimization is done, the updated \(\textbf{d}^\star\) is acquired. As a final step, QuadriFlow recomputes \( \textbf{p}_i^* \) and \( \textbf{p}_j^* \) based on \(\textbf{d}^\star\), yielding \( \textbf{p}_i^\star \) and \( \textbf{p}_j^\star \), the minimizers to the target constrained problem.

To achieve boundary alignment, we first identify the integer offsets that are associated with boundary vertices. Let \( \textbf{d}_{ij}^B \) denote such a case, which implies that either \( \textbf{v}_i \) or \( \textbf{v}_j \) is a boundary vertex. Local lattices of \( \textbf{v}_i \) and \( \textbf{v}_j \) are intended to align with the boundary. As a result, one axis of their local frames must coincide with the boundary tangent. Therefore, the 2D integer vector \( \textbf{d}_{ij}^B \) can only have three cases: \((0,0)\), \((m,0)\), and \((0,n)\), where \(m\) and \(n\) are non-zero integers. It means that \( \textbf{d}_{ij}^B \) is either fixed or an offset along the boundary. 

When solving the constrained optimization problem in Eqs. (\ref{eqmind}-\ref{eq11}) with also the boundary treatment, modification of \( \textbf{d}_{ij}^B \) needs a special attention. In principle, the zero component should always remain zero. More specifically, when \( \textbf{d}_{ij}^B \) = \((0,0)\), it is fixed during the optimization and we do not modify it. Second, when \( \textbf{d}_{ij}^B \) = \((m,0)\), the first component is allowed to change, whereas the second component is fixed to be zero. A similar treatment applies to the case when \( \textbf{d}_{ij}^B \) = \((0,n)\). These constraints on the modification of \( \textbf{d}_{ij}^B \) will ensure that the position field near the boundary is singularity free, thus yielding regular quad meshes along the boundary.

\begin{algorithm}
\caption{Minimization of the smoothness energy in the position field}\label{algo2}
\begin{algorithmic}[1]
\State Initialize the lattice origins    $\{\textbf{p}_i\}$ randomly;
\For{$\mathrm{iter} = 0$ \textbf{to} $\mathrm{max}\_\mathrm{iteration} $}
\For{$\mathrm{every \ vertex} \ \textbf{v}_i $}
\State $\mathrm{Find \  its \ adjacent }$ $\textbf{v}_j$;
    \State $\textbf{t}_{ij} = {\arg\min}\ \Vert \Gamma(\textbf{p}_i,\textbf{n}_i,\textbf{o}^*_i,\textbf{t})- \Gamma(\textbf{p}_j,\textbf{n}_j,\textbf{o}^*_j,\textbf{t})\Vert$;
    \State $\textbf{p}_i \gets \textbf{p}_i +\sum\limits_{j\in \textit{N}_i}\omega_{ij}\ \Gamma(\textbf{p}_{ji},\textbf{n}_i,\textbf{o}^*_i,\textbf{t}_{ij})$;
    \State $\textbf{p}_i \gets \mathrm{round}(\textbf{p}_i / \substack{\sum_{j\in \textit{N}_i}} \omega_{ij})$;
    \If{$\textbf{v}_i$ is a boundary vertex}
    \State Impose boundary constraints on $\textbf{p}_i$;
    \EndIf
    \EndFor
\EndFor
\For{$\mathrm{every \ triangular \ face} \ \Delta_{ijk}$}
    \State Calculate $\textbf{d}_{ij}$, $\textbf{d}_{jk}$, $\textbf{d}_{ki}$;
    \For{$\mathrm{every \ edge \ (p,q) \ of} \ \Delta_{ijk}$}
\If{$\textbf{v}_p \ \mathrm{or} \  \textbf{v}_q \ \mathrm{is \ a \ boundary \ vertex} $}
\If{$\textbf{d}_{pq} = (0,0)$}
\State Fix both components of $\textbf{d}_{pq}$;
\ElsIf{$\textbf{d}_{pq} = (m,0)$}
\State Fix the second component of $\textbf{d}_{pq}$;
\ElsIf{$\textbf{d}_{pq} = (0,n)$}
\State Fix the first component of $\textbf{d}_{pq}$;
\EndIf
\EndIf
\EndFor
    \State Modify free components of $\textbf{d}_{ij}$, $\textbf{d}_{jk}$, $\textbf{d}_{ki}$ to impose the regularity constraint;
     \State Modify free components of $\textbf{d}_{ij}$, $\textbf{d}_{jk}$, $\textbf{d}_{ki}$ to impose the consistency constraint;
\EndFor
\For{$\mathrm{every \ optimal} \ \textbf{p}_i^* \ \mathrm{from \ unconstrained \ optimization}$}
    \If{$\textbf{v}_i$ is a boundary vertex}
    \State Impose boundary constraints on $\textbf{p}_i^*$;
    \EndIf
    \State Recompute to obtain updated $\textbf{p}_i^\star$ based on \(\textbf{d}^\star\) (from Lines 26 and 27);
\EndFor
\end{algorithmic}
\end{algorithm}

\begin{figure}[h]
  \subcaptionbox{Input triangle mesh}[0.28\textwidth]{\includegraphics[width=1.05\linewidth]{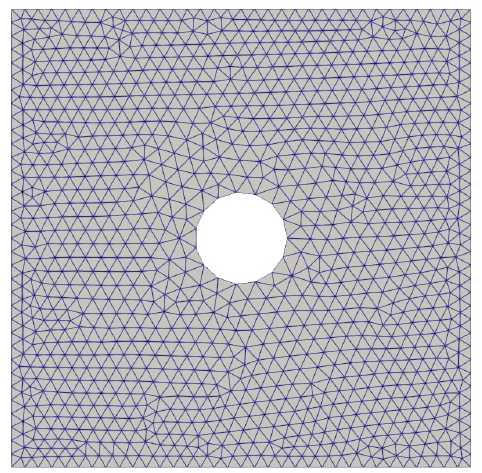}}
  \subcaptionbox{Without boundary constraints}[0.37\textwidth]{\includegraphics[width=0.8\linewidth]{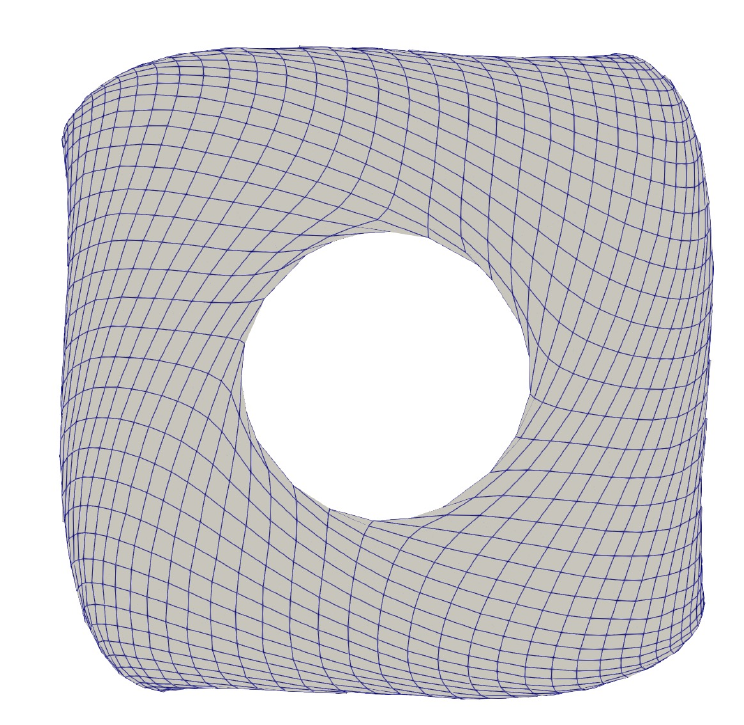}}
  \subcaptionbox{With boundary constraints}[0.33\textwidth]{\includegraphics[width=0.9\linewidth]{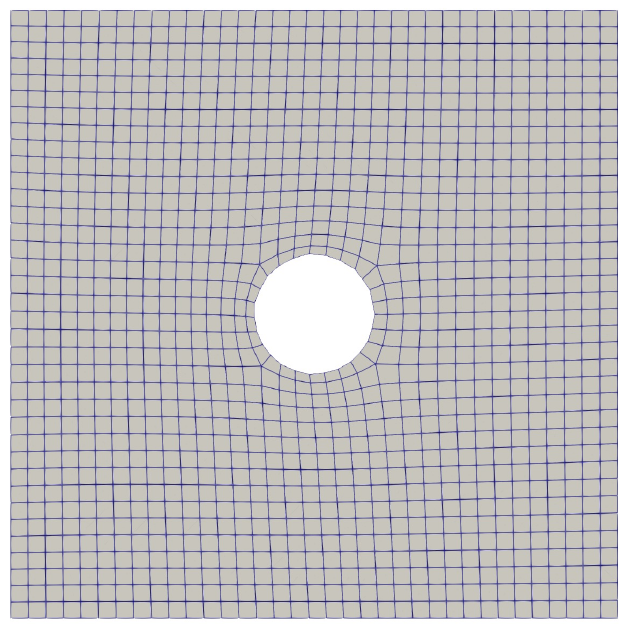}}
  \caption{Demonstration of the necessity of imposing boundary constraints. (a) The input triangle mesh for the plate-with-a-hole model, (b) the quad mesh generated by the original QuadriFlow without boundary treatment, and (c) the quad mesh generated after imposing boundary constraints.}
  \label{comparison}
\end{figure}

Once boundary constraints are added, a valid quad mesh for an open surface can be obtained. Fig.~\ref{comparison} shows a comparative study on a plate-with-a-hole model, underscoring the necessity of boundary constraints. The initial triangle mesh, shown in Fig.~\ref{comparison}(a), serves as the input for quad meshing. Fig.~\ref{comparison}(b) presents the quad mesh generated by the original QuadriFlow that does not have the boundary treatment. It exhibits noticeable discrepancies from the input geometry along the boundaries. 
In contrast, the quad mesh generated with the boundary treatment is shown in Fig.~\ref{comparison}(c), which accurately aligns with the boundaries. Moreover, because of the constraint imposed on \( \textbf{d}_{ij}^B \), boundary points are all regular in the sense that they correspond to either an ``L"-junction or a ``T"-junction; see the precise definition in the following.

% \begin{figure}[htbp]
% 	\centering
% 	\begin{subfigure}{0.31\linewidth}
% 		\centering
% 		\includegraphics[width=1\linewidth]{comparison_input.pdf}
%             \caption{Input triangle mesh}
% 	\end{subfigure}
% 	\centering
% 	\begin{subfigure}{0.33\linewidth}
% 		\centering\includegraphics[width=1\linewidth]{without_boundary_constraints.pdf}
%             \caption{{Without boundary constraints}}
% 	\end{subfigure}
% 	\centering
% 	\begin{subfigure}{0.31\linewidth}
% 		\centering
% 		\includegraphics[width=1\linewidth]{with_boundary_constraints.pdf}
%             \centering
%             \caption{With boundary constraints} 
% 	\end{subfigure}
%         \caption{Demonstration of the necessity of imposing boundary constraints. (a) The input triangle mesh for the plate-with-a-hole model, (b) the quad mesh generated by the original QuadriFlow without boundary treatment, and (c) the quad mesh generated after imposing boundary constraints.}
%         \label{comparison}
% \end{figure}

\section{Patch extraction and simplification}\label{simplify}
Once the quad mesh is ready, we proceed to extract \emph{patches} out of it. A patch is simply a four-sided regular grid. It can be parameterized in a straightforward way through the classical Floater's algorithm~\cite{ref:floater97, ref:floater03}, such that later we can easily fit a (non-trimmed) NURBS surface to it with a user-controllable tolerance. Note that the quad mesh itself is usually dense and not suitable to directly serve as a control mesh. Otherwise, we end up with an unnecessarily high-resolution spline representation, which, however, may still possess a considerable geometric error from the original model because of the non-interpolatory nature of splines. Some other work appeals to Coons patches for the reconstruction purpose~\cite{ref:hiemstra20}, but such patches can only capture patch boundaries and have no control over the patch interior, hence possibly introducing a large fitting error. 

The key to patch extraction is to find the separatrices, i.e., the line segments that deliminate different rectangular patches. Finding them is rather straightforward for open quad meshes. To start with, we introduce a few terminologies to facilitate the following discussion. An \emph{extraordinary point} (EP) is an interior point shared by other than four edges, or a boundary point shared by more than three edges; otherwise it is a regular point. In QuadriFlow-based quad meshing, an EP corresponds to a singularity in the orientation field, as the position fields is singularity free. Recall that a \emph{corner} is a boundary point shared by two edges, or a boundary point whose adjacent boundary edges have a sharp turn in directions (determined by a user-specified parameter). A separatrix starts from an extraordinary point or a corner and traverses along the same consistent direction until it hits another extraordinary point, corner, or a boundary. A sharp feature line (e.g., creases) also corresponds to a separatrix. The collection of all the separatrices partitions the whole quad mesh into a set of conforming rectangular patches.

\begin{figure}[h]
  \subcaptionbox{Singularity clustering}[0.29\textwidth]{\includegraphics[width=0.95\linewidth]{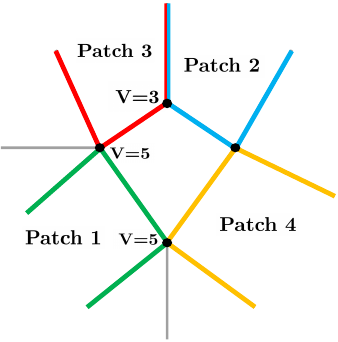}}
  \hfill
  \subcaptionbox{Singularity misalignment}[0.5\textwidth]{\includegraphics[width=1\linewidth]{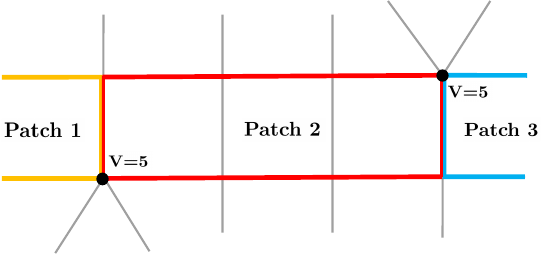}}
  \caption{Two sources that lead to tiny and slender patches: singularity clustering and singularity misalignment. (a) Singularity clustering is the case when multiple extraordinary points appear in a single element. (b) Singularity misalignment is the case when two extraordinary points appear at the opposite corners of a patch.}
  \label{ps}
\end{figure}

The placement of EPs plays a crucial role in patch extraction. While the field-aligned parameterization in QuadriFlow tends to produce well-structured quad meshes, the placement of singularities is usually resolved under several competing constraints. It is not guaranteed to be optimal in general, which may lead to tiny or slender patches. These patches are aesthetically displeasing and redundant, motivating us to simplify the patch structure. There are two sources that lead to tiny/slender patches: singularity clustering and singularity misalignment, as shown in Fig.~\ref{ps}. We discuss their treatment one by one in the following.

% Slender patches often contain elements with a high aspect ratio. When used in simulation, deformation in such patches may immediately cause fold-over (i.e., negative Jacobian), rendering the simulation non-convergent. Moreover, 

% \begin{figure}[h]
%     \centering
%     \begin{subfigure}[b]{0.25\textwidth}
%            \centering
%            \includegraphics[width=1\textwidth]{singularity_clustering.pdf}
%             \caption{Singularity clustering}
%     \end{subfigure}
%     \hspace{1.5cm}
%     \begin{subfigure}[b]{0.5\textwidth}
%             \centering
%             \includegraphics[width=1\textwidth]{singularity_misalignment.pdf}
%             \caption{Singularity misalignment}
%     \end{subfigure}
%     \caption{Two sources that lead to tiny and slender patches: singularity clustering and singularity misalignment. (a) Singularity clustering is the case when multiple extraordinary points appear in a single element. (b) Singularity misalignment is the case when two extraordinary points appear at the opposite corners of a patch.}
%     \label{ps}
% \end{figure}

\subsection{Treatment of singularity clustering}\label{subsecsc}
Singularity clustering corresponds to the case when multiple EPs appear in a single element; see Fig.~\ref{ps}(a). This case occurs when QuadriFlow is unable to yield a smooth orientation field in a small local region that involves an unnecessarily dense triangle mesh. Each vertex of an element may or may not be an EP, and when it is an EP, it may have a different \emph{valence}, i.e., the number of edges sharing this point. This renders infinitely many cases and makes them challenging to solve. Fortunately, based on various tests on the large dataset~\cite{ref:quadriflow, ref:jakob15, chang2015shapenet}, QuadriFlow tends to only generate two kinds of extraordinary points: valence-3 and valence-5. Moreover, we perform a statistical analysis of over 1,000 different inputs to check the case of clustering, in which we find that a single element may involve only two EPs or three EPs but no case of four EPs. In an element with two EPs, the two EPs are right next to each other (the diagonally opposite case will be discussed in Sec~\ref{subsecsm}). Depending on the valances of the EPs, this case can be further divided into three sub-cases: 3-3, 3-5, and 5-5. For an element with three EPs, the only clustering patterns that appear in the above-mentioned tests are 3-3-5 and 3-5-5. Among them, the most frequently occurring cases are 3-5, 3-3-5, and 3-5-5.

\begin{figure}[h]%
\centering
\includegraphics[width=1\textwidth]{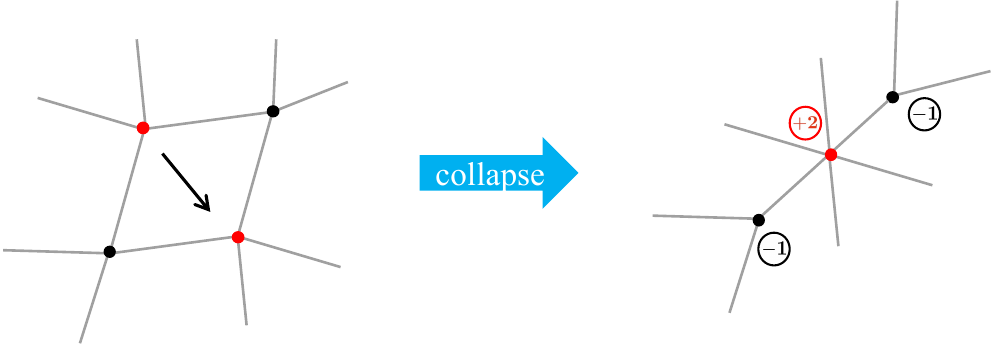}
\caption{The diagonal collapsing strategy in~\cite{feng2021q}: by collapsing the two opposite vertices of valence 4 (red points), the valence of the collapsed vertex increases by two and becomes six, whereas the valence of the non-collapsed vertices (black points) decreases by one and becomes three.}\label{diacol}
\end{figure}

To address the issue of singularity clustering, we propose a method that is based on the collapsing idea proposed in~\cite{feng2021q}. As a mesh simplification scheme, it introduced a strategy to collapse the two opposite vertices of an element into a single vertex. In doing so, the valence of the collapsed vertex increases by two, whereas the valence of each non-collapsed vertices decreases by one, as illustrated in Fig.~\ref{diacol}. 

Motivated by the diagonal collapsing idea, we propose a dedicated collapsing method for adjacent vertices of an element to deal with singularity clustering. Let us consider two adjacent vertices of valences \(\alpha\) and \(\beta\), with \(\alpha \in \mathbb{N} \) and \(\beta \in \mathbb{N} \). By collapsing these two vertices, the resulting collapsed vertex has a valence \(\alpha+\beta-4\). In the case of two regular vertices (i.e., valence 4), the collapsing process yields again a regular vertex and does not introduce new EPs.

\begin{figure}[h]
  \subcaptionbox{Treatment of Case 3-5}[1\textwidth]{\includegraphics[width=0.95\linewidth]{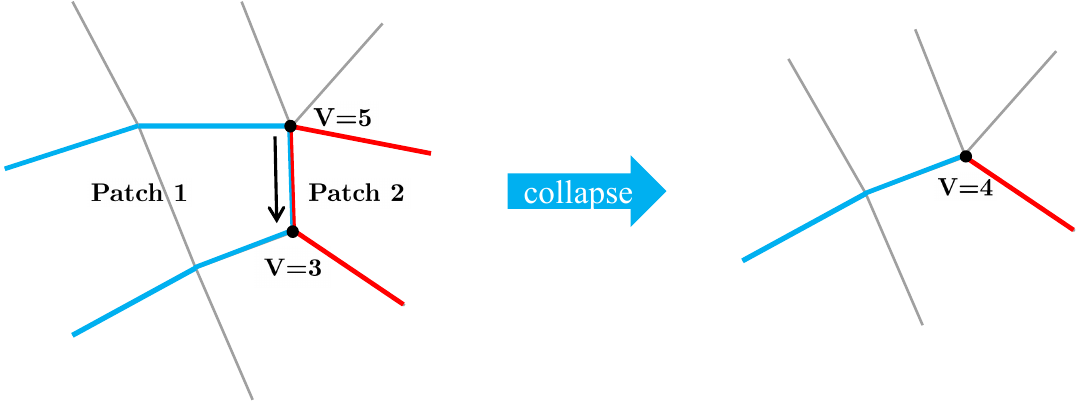}}
  \subcaptionbox{Treatment of Case 5-5}[1\textwidth]{\includegraphics[width=1\linewidth]{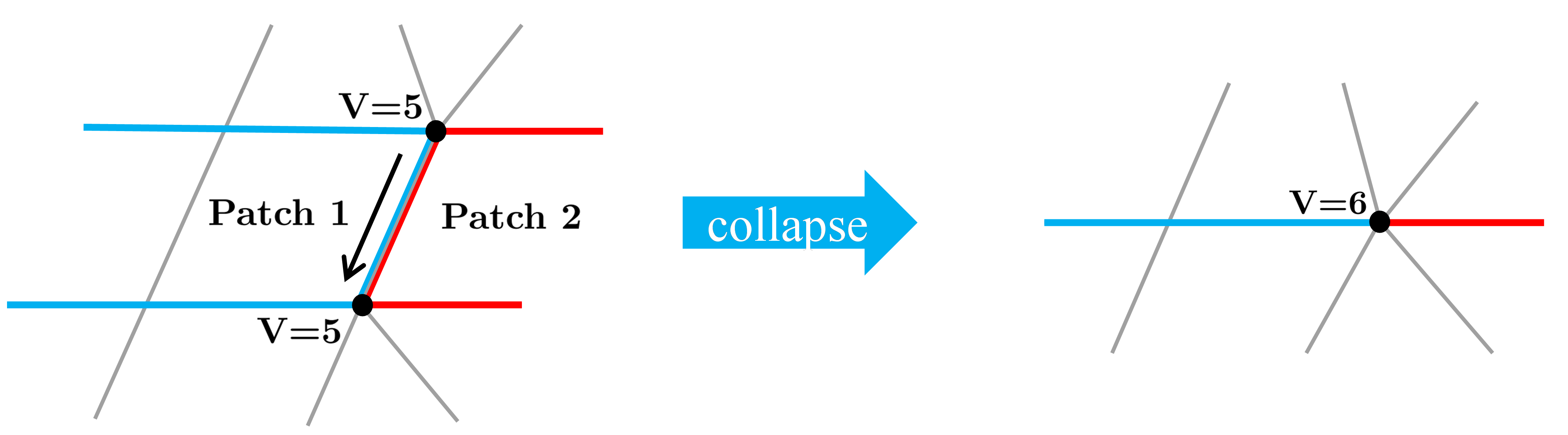}}
  \caption{Collapsing of two adjacent EPs to deal with singularity clustering. (a) Treatment of Case 3-5: collapsing the valence-5 and valence-3 EPs into a regular vertex. (b) Treatment of Case 5-5: collapsing the two valence-5 EPs into a valence-6 EP.}
  \label{35}
\end{figure}

We then apply this strategy to dealing with singularity clustering. First, for a two-EP element, Cases 3-5 and 5-5 are illustrated in Fig.~\ref{35}(a) and Fig.~\ref{35}(b), respectively. In Case 3-5, the two EPs become a regular vertex after collapsing, whereas in Case 5-5, the two EPs become a valence-6 vertex. Both cases lead to an effective reduction of EPs. However, Case 3-3 cannot be handled by this collapsing idea, because it will yield an invalid valence-2 vertex in the interior mesh. On the other hand, this case is rarely encountered in practical applications, so we postpone its treatment in the future.

\begin{figure}[h]
  \subcaptionbox{Treatment of Case 3-5-5}[1\textwidth]{\includegraphics[width=\linewidth]{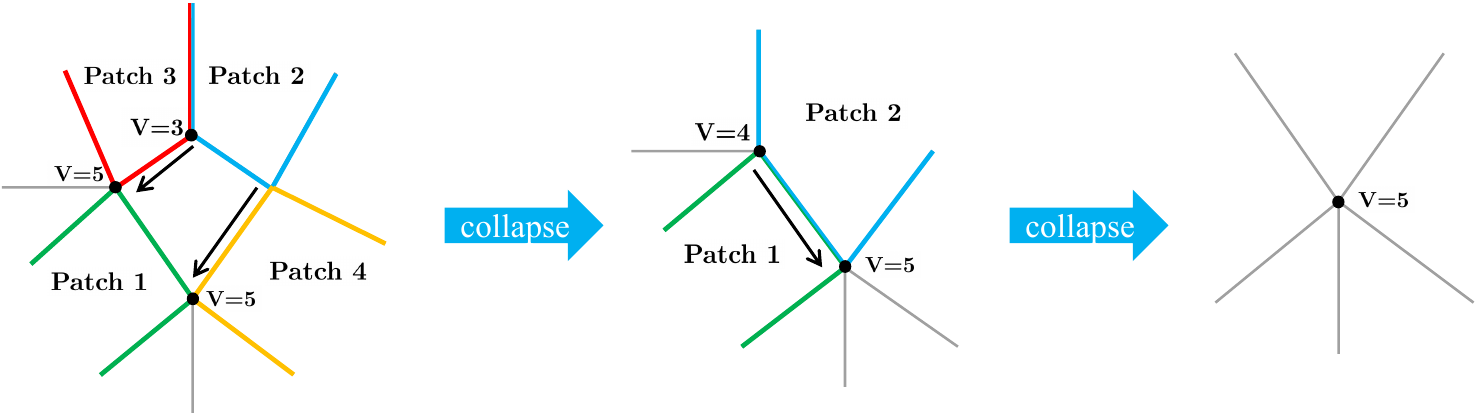}}
  \subcaptionbox{Treatment of Case 3-3-5}[1\textwidth]{\includegraphics[width=1\linewidth]{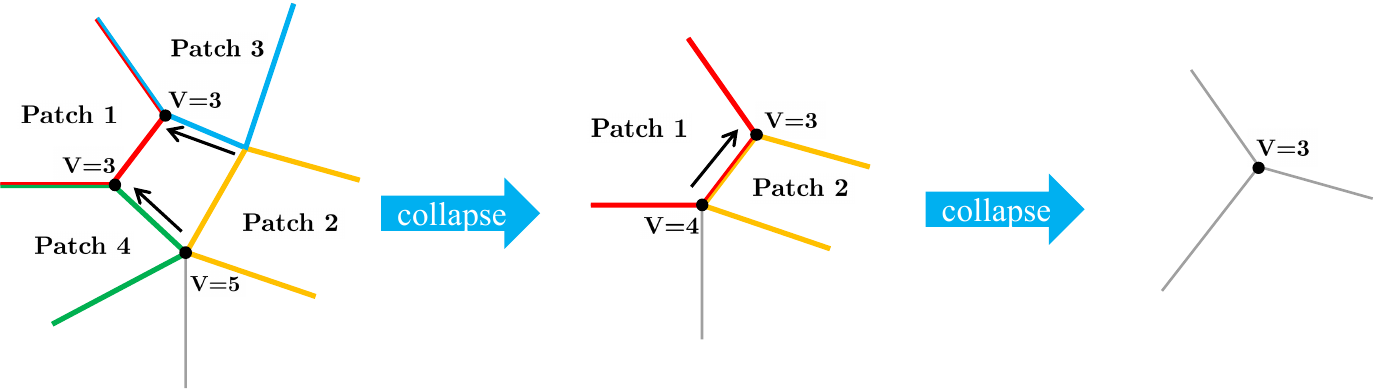}}
  \caption{Collapsing of two adjacent EPs to deal with singularity clustering. (a) Treatment of Case 3-5: collapsing the valence-5 and valence-3 EPs into a regular vertex. (b) Treatment of Case 5-5: collapsing the two valence-5 EPs into a valence-6 EP.}
  \label{355}
\end{figure}

Second, for an element with three EPs, collapsing for Cases 3-3-5 and 3-5-5 is shown in Fig.~\ref{355}(a) and Fig.~\ref{355}(b), respectively. This is achieved by applying collapsing twice: the first time by collapsing the valence-3 and valence-5 vertices into a regular vertex, and the second time along the other direction. Note that according to the proposed rule, collapsing a regular vertex with any EP results in a vertex of valence identical to the EP. Therefore, collapsing always ensures a reduction in the number of EPs. Treatment of the multi-EP elements not only removes their corresponding tiny patches, but also results in a global operation that eliminates all the involved slender patches throughout the quad mesh; see Figs.~\ref{35} and~\ref{355} for illustration.

% \begin{figure}[h]
%     \centering
%         \begin{subfigure}[b]{1\textwidth}
%            \centering
%            \includegraphics[width=0.95\textwidth]{35.pdf}
%             \caption{Treatment of Case 3-5}
%     \end{subfigure}
%     \begin{subfigure}[b]{1\textwidth}
%            \centering
%            \includegraphics[width=1\textwidth]{55.png}
%             \caption{Treatment of Case 5-5}
%     \end{subfigure}
%  \caption{Collapsing of two adjacent EPs to deal with singularity clustering. (a) Treatment of Case 3-5: collapsing the valence-5 and valence-3 EPs into a regular vertex. (b) Treatment of Case 5-5: collapsing the two valence-5 EPs into a valence-6 EP.}
%     \label{35}
% \end{figure}

%\begin{figure}[h]%
%\centering
%\includegraphics[width=1\textwidth]{35.pdf}
%\caption{Treatment of the 3-5 clustering: coalesce the valence-5 and valence-3 points into a regular point.
%}\label{35}
%\end{figure}
% \begin{figure}[h]
%     \centering
%     \begin{subfigure}[b]{1\textwidth}
%            \centering
%            \includegraphics[width=1\textwidth]{355.pdf}
%             \caption{Treatment of Case 3-5-5}
%     \end{subfigure}
%     \begin{subfigure}[b]{1\textwidth}
%             \centering
%             \includegraphics[width=1\textwidth]{335.pdf}
%             \caption{Treatment of Case 3-3-5}
%     \end{subfigure}
%     \caption{Collapsing of three EPs in a single element to deal with singularity clustering. (a) Treatment of Case 3-5-5, leading to a single valence-5 EP. (b) Treatment of Case 3-3-5, leading to a single valence-3 EP.}
%     \label{355}
% \end{figure}

\subsection{Treatment of singularity misalignment}\label{subsecsm}

Singularity misalignment corresponds to the case when two EPs appear at the opposite corners of a patch, rather than being connected by the same separatrix; see Fig.~\ref{ps}(b). This problem arises because it is inherently difficult for QuadriFlow, a method based on local operators, to enforce global constraints on the placement of EPs during the optimization. As a result, singularity misalignment is a prevalent issue in the resulting quad meshes.

\begin{figure}[h]%
\centering
\includegraphics[width=1\textwidth]{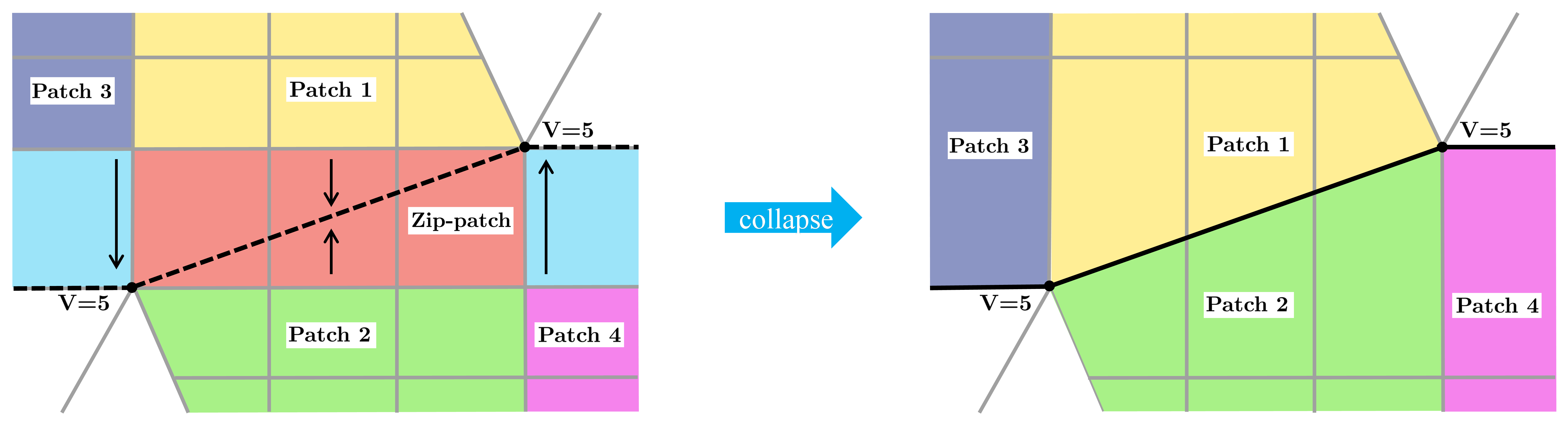}
\caption{Collapsing towards the diagonal of a zip-patch to deal with singularity misalignment.}
\label{sms}
\end{figure}

The treatment of singularity misalignment is primarily based on the simplification idea in~\cite{viertel2019coarse}, on top of which we extend it to support sharp features that often appear in mechanical parts. After patch extraction, the so-called \textit{zip-patches} are identified when two corners of a patch are EPs and they are opposite to one another; see Fig.~\ref{sms}. Such a patch can be eliminated by collapsing towards the diagonal that connects the two EPs. Note that the valences of the two EPs remain the same during collapsing. Subsequently, the neighboring patches are also collapsed, which propagates through the entire mesh and thus eliminates the involved slender patches.

During patch extraction, sharp features are taken into account and they serve as separatrices that separate different patches. A sharp feature consists of a sequence of consistently connected edges. A sharp edge is identified when the outward normals of its adjacent faces form an angle exceeding a user-defined threshold (e.g., \(\pi/3\)). Sharp edges are restricted to move only along their tangential direction. As a consequence, if the long side of a slender patch is identified as a sharp feature, collapsing of this side towards the diagonal is not allowed; otherwise the sharp feature will move off its tangent, leading to a significant increase in the fitting error. On the other hand, the short side of a slender patch always moves along its tangential direction during collapsing, and thus collapsing of the short side is allowed.

Once the simplified patch layout is obtained, we finish the whole workflow with a step of standard spline fitting (see details in Appendix~\ref{app:fitting}). To this end, we reparameterize the input CAD model entirely as a watertight spline representation.

\section{Numerical examples}\label{result}
In this section, we demonstrate the efficacy and efficiency of the proposed pipeline through several engineering CAD models. The entire pipeline is implemented in C++. All examples are run on a PC with the 12th Gen Intel(R) Core(TM) i7-12700 CPU and 32GB RAM.

\begin{table}[b]
\caption{Statistics of all the tested models.}\label{table1}
\begin{tabular*}{\textwidth}{@{\extracolsep\fill}llllllll}
\toprule
Model & TF & TM(s) & MR(s) & QM(s) & PS(s) & NF(s) & Total(s) \\
\midrule
Plane plate & 112,822 & 5.9 & 2.1  & 16.9 & 1.3 & 1.3 & 27.5\\
Ship head & 235,140 & 31.5  & 10.4  & 48.9  & 0 & 0.9 & 91.7\\
Car body & 235,651 & 28.9  & 8.9 & 21.6 & 1.3 & 1.9 & 62.6\\
Plane head & 16,893,754 & 643.3  & 0 & 901.4 & 1.6 & 3.8 & 1550.1\\
Ship hull & 11,375,654 & 423.4  & 243.7 & 552.4 & 1.6 & 19.1 & 1240.2\\
\bottomrule % 使用bottomrule代替botrule
\end{tabular*}
\footnotetext{TF: number of triangular faces; TM: triangular meshing; MR: mesh repair; QM: quad meshing; PS: patch simlification; NF: NURBS fitting. }
\end{table}

The overall statistics of all the tested models are summarized in Table~\ref{table1}. We test our pipeline on five models: a plane plate, a ship head, a car body, a plane head, and a ship hull. The scale of the input triangle meshes and the time cost at each key step are reported, including the number of faces in triangle meshes, time for triangle mesh generation (by Gmsh), time for mesh repair, time for quad meshing, time for patch simplification, and time for NURBS fitting. All the models are reconstructed as multi-patch NURBS of degree 2 or 3. We observe that the largest model (i.e., the plane head) has more than 10 million triangular faces and the whole pipeline is finished around 25 mins. We further observe that among all steps, triangulation and quad meshing dominate the running time, whereas the overhead for patch simplification and surface fitting is mostly negligible. More importantly, the time for mesh repair and quad meshing roughly increases linearly with the size of triangle meshes, demonstrating the scalability of the proposed pipeline (see more results later). In what follows, we discuss the reconstruction of these models in detail.

\begin{figure}[htb]
    \centering
    \begin{subfigure}[b]{0.45\textwidth}
           \centering
           \includegraphics[width=0.58\textwidth]{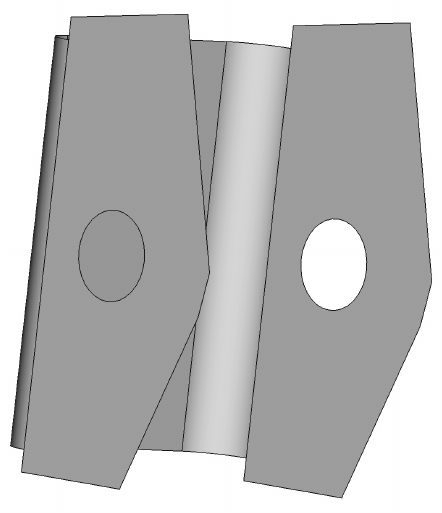}
            \caption{Trimmed CAD model}
    \end{subfigure}
    \hfill
    \begin{subfigure}[b]{0.45\textwidth}
            \centering
            \includegraphics[width=0.75\textwidth]{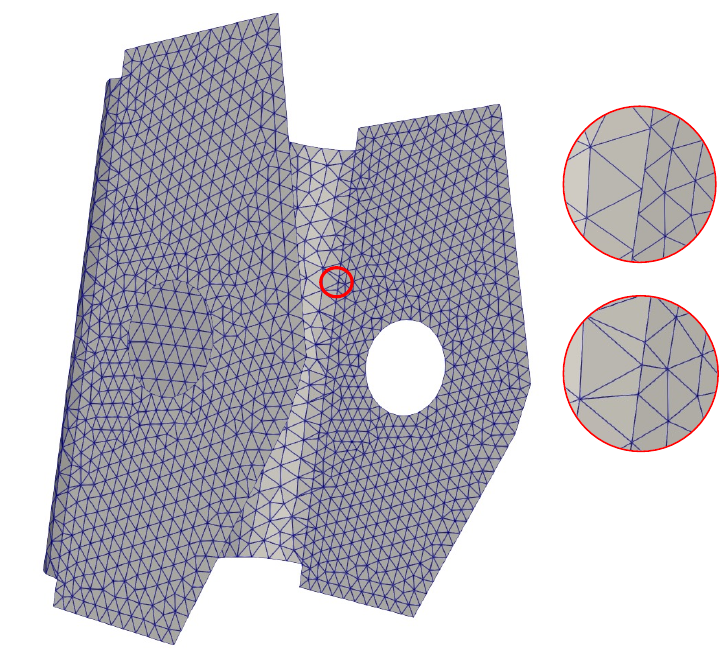}
            \caption{Triangulation and mesh repair}
    \end{subfigure}
    \begin{subfigure}[b]{0.45\textwidth}
           \centering
           \includegraphics[width=0.58\textwidth]{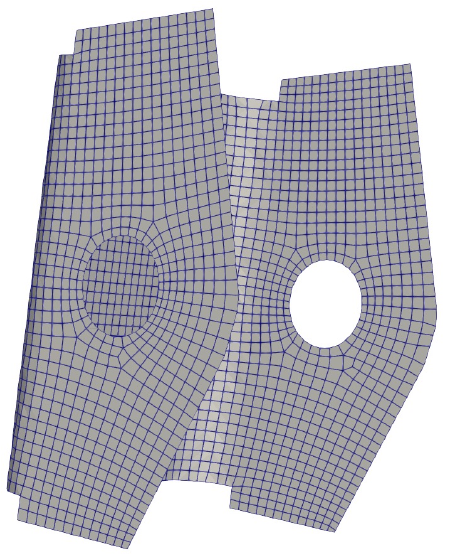}
            \caption{Quad mesh generation}
    \end{subfigure}
    \hfill
    \begin{subfigure}[b]{0.45\textwidth}
            \centering
            \includegraphics[width=0.6\textwidth]{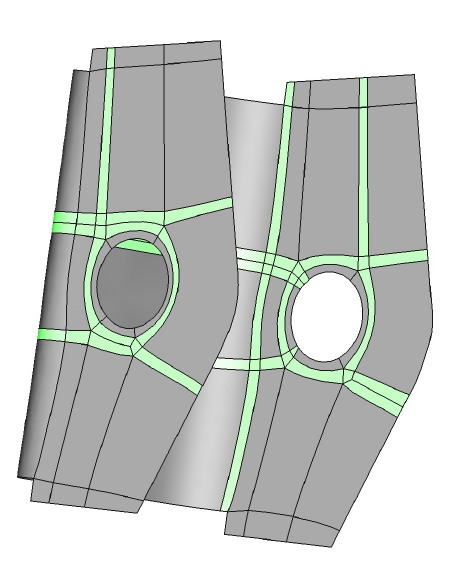}
            \caption{Patch extraction}
    \end{subfigure}
    \begin{subfigure}[b]{0.45\textwidth}
            \centering
            \includegraphics[width=0.65\textwidth]{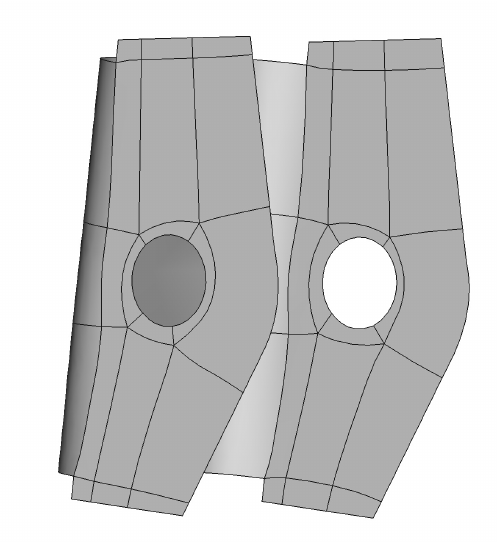}
            \caption{Patch simplification and fitting}
    \end{subfigure}
    \caption{Reconstruction of a plane plate model as a watertight muti-patch NURBS surface.}\label{fig11}
\end{figure}

\subsection{Reparameterization}
Fig.~\ref{fig11} shows the reconstruction of a plane plate model. The input CAD model is an open surface with sharp corners, where the two holes are created by trimming. As shown in Fig.~\ref{fig11}(a), it is composed of five patches, each of which is bounded by black lines. The topology information of this model is problematic, namely, the patches are isolated surfaces rather than being properly connected. On the other hand, the geometric discrepancy at the patch interfaces is indistinguishable. Such a wrong-topology-but-``correct"-geometry pattern leads to invisible gaps and hinders further editing of the model. This is often encountered in engineering applications due to the model transfer/conversion between different CAD systems. With such a CAD model as input, a triangle mesh is generated using open-source software Gmsh or OpenCASCADE, where Gmsh is preferred as it yields much more uniform meshes than the latter. Due to the topological issue in the CAD model, gaps immediately appear in the triangle mesh; see Fig.~\ref{fig11}(b). It involves two cases from the mesh point of view: duplicated points and points lying on edges. A simple repair step is followed to automatically rebuild the correct mesh connectivity; see Appendix~\ref{app:fixing} for details. At this point, the geometric model, now approximated by a triangle mesh within a user-controllable tolerance, becomes watertight, paving the way for the subsequent step of quad meshing. This intermediate step of triangulation is necessary for an untrimming pipeline as almost all the methods of quad meshing rely on triangle meshes as input.

\begin{figure}[hb]
    \centering
    \begin{subfigure}[b]{0.45\textwidth}
           \centering
           \includegraphics[width=0.85\textwidth]{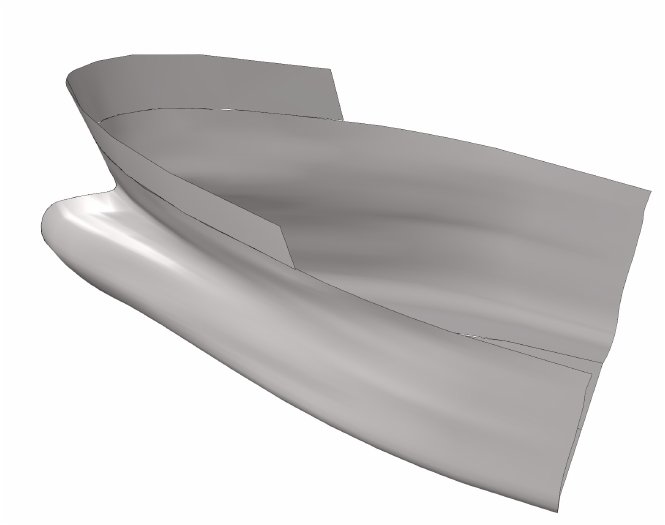}
            \caption{Trimmed CAD model}
    \end{subfigure}
    \hfill
    \begin{subfigure}[b]{0.45\textwidth}
            \centering
            \includegraphics[width=0.85\textwidth]{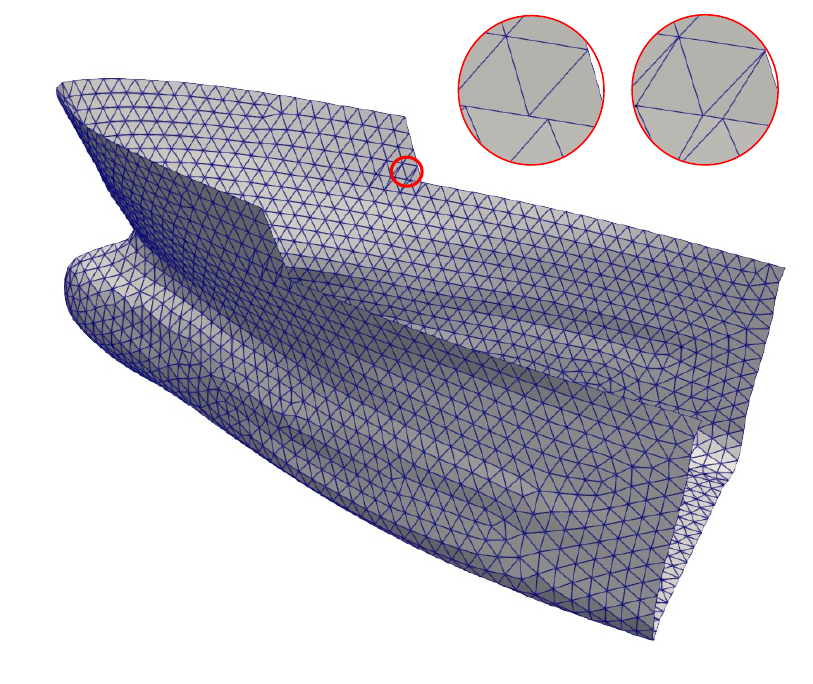}
            \caption{Triangulation and mesh repair}
    \end{subfigure}
    \begin{subfigure}[b]{0.45\textwidth}
           \centering
           \includegraphics[width=0.8\textwidth]{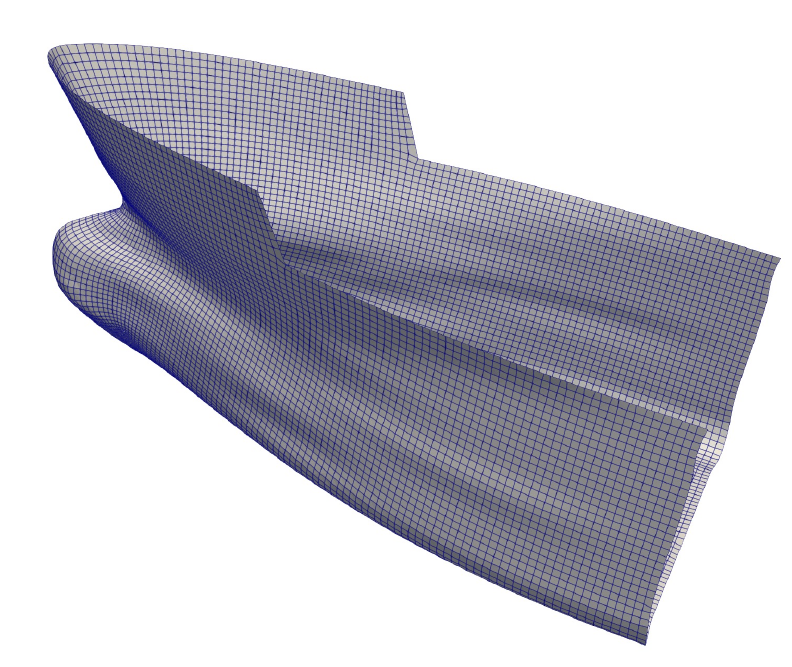}
            \caption{Quad mesh generation}
    \end{subfigure}
    \hfill
    \begin{subfigure}[b]{0.45\textwidth}
            \centering
            \includegraphics[width=0.85\textwidth]{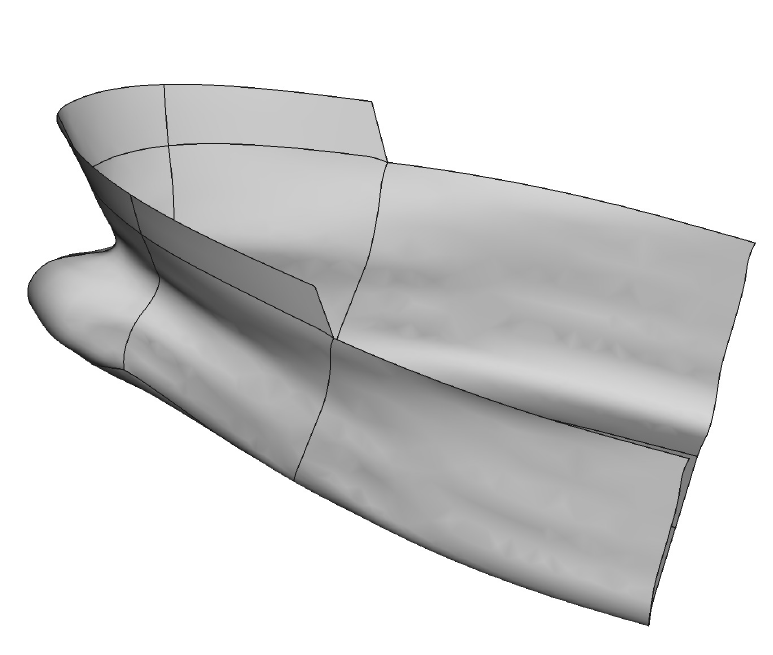}
            \caption{Patch extraction and fitting}
    \end{subfigure}
    \caption{Reconstruction of a ship head model as a watertight muti-patch NURBS surface.}\label{fig12}
\end{figure}

With the watertight triangle mesh as input, we apply our enhanced version of QuadriFlow to generate a corresponding quad mesh. The meshing process is fast and fully automatic once a key parameter, called a magnitude factor \(\gamma\), is properly set. The parameter \(\gamma\) indicates the ratio between the average edge length of the output quad mesh and that of the input triangle mesh. According to tests on various models with different resolutions, valid quad meshes (i.e., free of severe distortion and meaningful distribution of EPs) usually can be obtained when \(\gamma \in [1, 2]\). For simplicity, \(\gamma\) is set to be 1.4 unless stated otherwise. This value empirically strikes a good balance to achieve accuracy and the effective capture of geometric features at the same time. The resulting quad mesh preserves all the boundary features as expected; see Fig.~\ref{fig11}(c). 
%The resolution of the quad mesh needs to be smaller than that of the triangle mesh, so we require \(\gamma > 1\).

Once the quad mesh is ready, the patch extraction is straightforward. However, we observe in Fig.~\ref{fig11}(d) that slender patches appear due to singularity clustering and misalignment. Patch simplification is then performed to eliminate such patches, yielding a much simplified layout. It is quantified in terms of number of patches; see Table~\ref{table2} for the comparison before and after patch simplification, where the number of patches is reduced by more than half through simplification.

Finally, based on the quad mesh with a simplified layout, a NURBS surface is fit to each patch in a least-squares manner; see Appendix~\ref{app:fitting} for details. A multi-patch representation is obtained, where conforming interfaces are a straightforward result inherited from the quad mesh. The final result, i.e., a watertight multi-patch NURBS representation, is shown in Fig.~\ref{fig11}(e).

\begin{figure}[h]
    \centering
    \begin{subfigure}[b]{0.45\textwidth}
           \centering
           \includegraphics[width=0.86\textwidth]{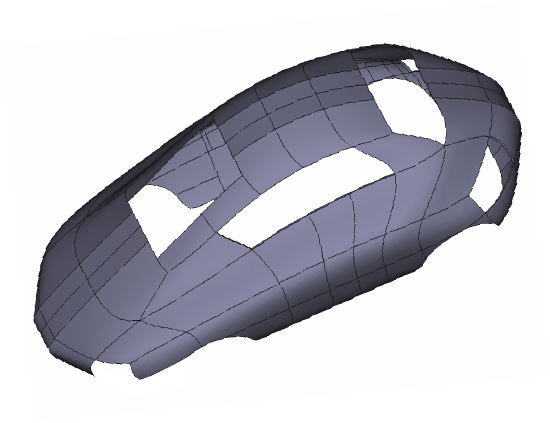}
            \caption{Trimmed CAD model}
    \end{subfigure}
    \hfill
    \begin{subfigure}[b]{0.45\textwidth}
            \centering
            \includegraphics[width=1\textwidth]{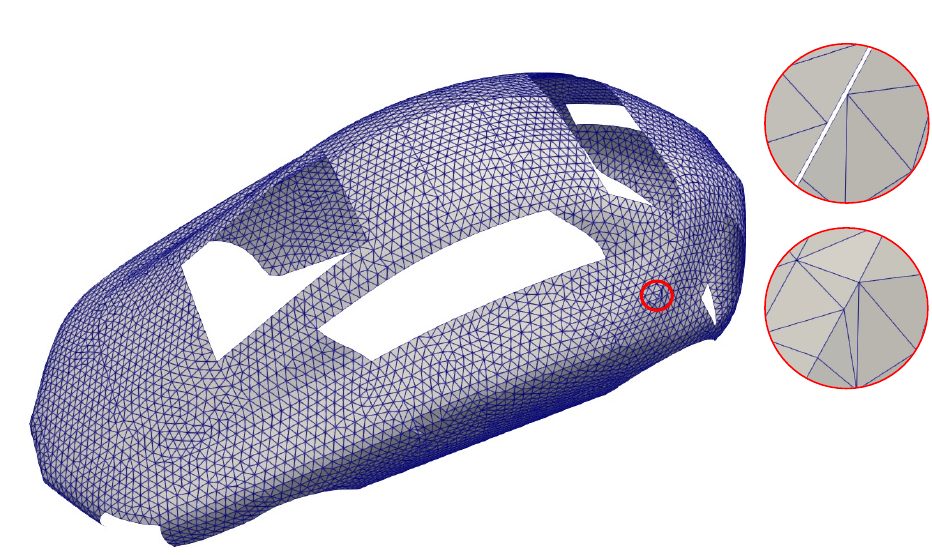}
            \caption{Triangulation and mesh repair}
    \end{subfigure}
    \begin{subfigure}[b]{0.45\textwidth}
           \centering
           \includegraphics[width=0.92\textwidth]{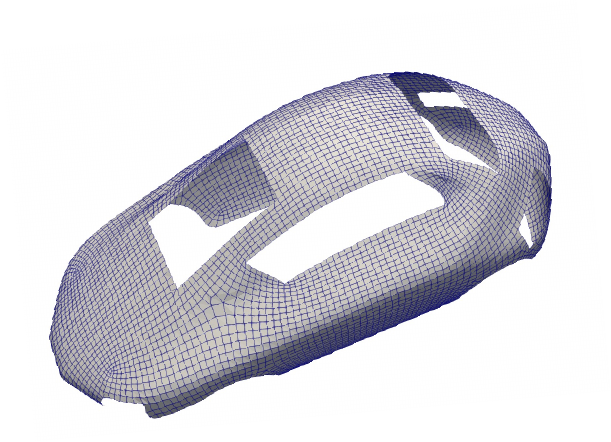}
            \caption{Quad mesh generation}
    \end{subfigure}
    \hfill
    \begin{subfigure}[b]{0.45\textwidth}
            \centering
            \includegraphics[width=0.82\textwidth]{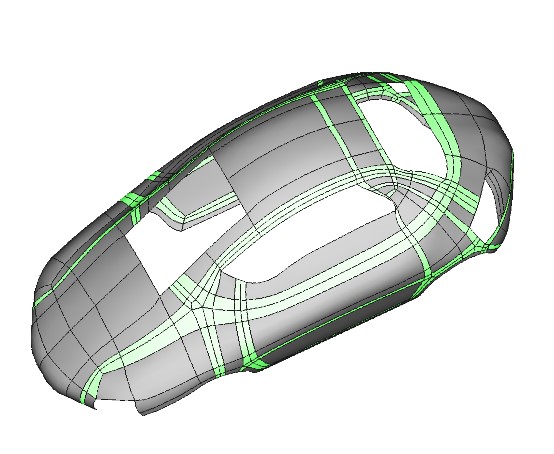}
            \caption{Patch extraction}
    \end{subfigure}
    \begin{subfigure}[b]{0.45\textwidth}
            \centering
            \includegraphics[width=0.8\textwidth]{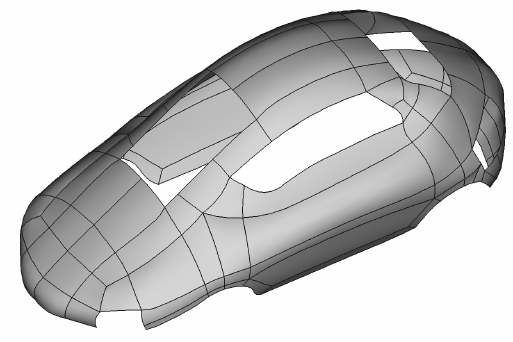}
            \caption{Patch simplification and fitting}
    \end{subfigure}
    \caption{Reconstruction of a car body model as a watertight muti-patch NURBS surface.}\label{fig13}
\end{figure}

The second model is a ship head from an online gallery of GrabCAD~\cite{ref:grabCAD}, which exhibits drastically varying curvature. Gaps also exist in the model. Following a similar procedure to the previous model, we can reconstruct it again as a watertight multi-patch NURBS; see Fig.~\ref{fig12}. The final layout is intuitively the best that one could hope for, which is symmetric and features the smallest number of patches while respecting all the geometric features, such as corners and high-curvature regions around the head. Note that in this model, patch simplification is not needed because the quad mesh obtained from QuadriFlow already has the simplest possible layout, where there is no issue of singularity clustering or misalignment. 
%Singularity positions are optimum, enabling the creation of a high-quality reconstruction without additional patch simplification. 

\begin{table}[h]
\caption{Number of patches before and after patch simplification}\label{table2}
\begin{tabular*}{1\textwidth}{@{\extracolsep\fill}lll}
\toprule
Model & Before & After  \\
\midrule
Plane plate & 85 & 39 \\
Ship head & 14 & 14 \\
Car body & 398 & 112 \\
Plane head & 28 & 14 \\
Ship hull & 87 & 34 \\
\bottomrule
\end{tabular*}
\end{table}

The third model is a car body with numerous internal boundaries and sharp corners. Due to the presence of large holes, the size of features in different areas varies drastically. The input CAD model is again a trimmed NURBS with gaps between adjacent patches. The proposed pipeline is then applied to the model for an automatic reconstruction. Note that there exists a large number of tiny/slender patches after performing patch extraction directly from the quad mesh. Patch simplification, on the other hand, can effectively eliminate all such patches and thus significantly reduce the number of patches, from 398 to 112; see Table~\ref{table2}. Eventually, a watertight representation with a clean layout is achieved; see Fig.~\ref{fig13}.

\begin{figure}[h]
    \centering
    \begin{subfigure}[b]{0.45\textwidth}
           \centering
           \includegraphics[width=0.75\textwidth]{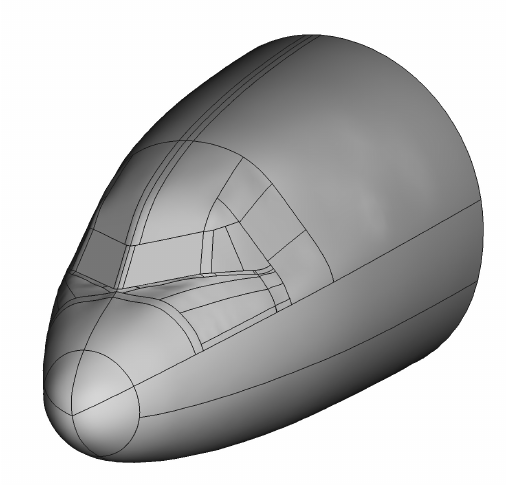}
            \caption{Trimmed CAD model}
    \end{subfigure}
    \hfill
    \begin{subfigure}[b]{0.45\textwidth}
            \centering
            \includegraphics[width=0.9\textwidth]{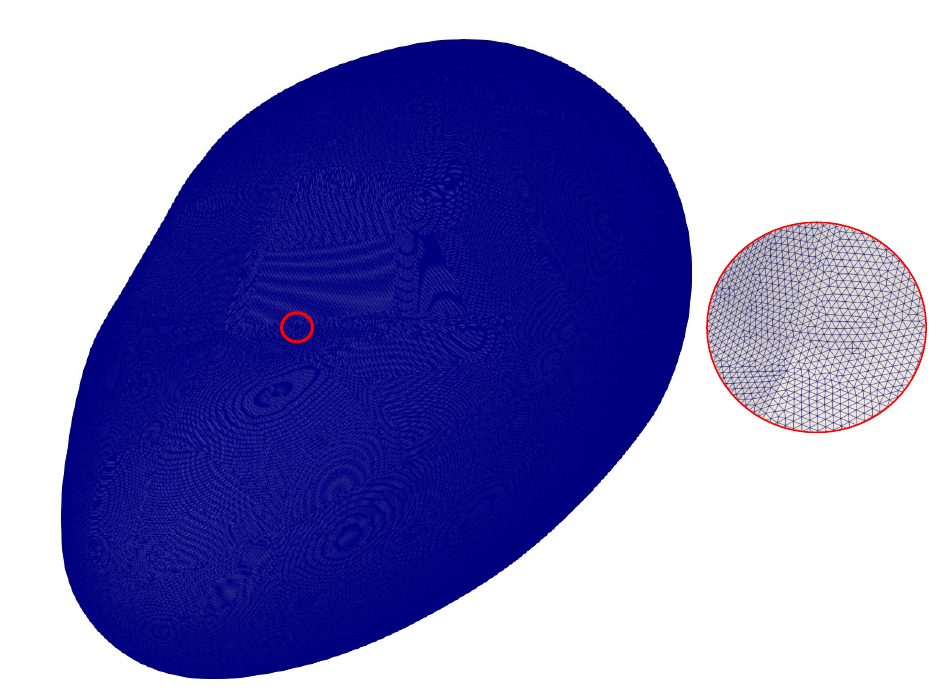}
            \caption{Triangulation}
    \end{subfigure}
    \begin{subfigure}[b]{0.45\textwidth}
           \centering
           \includegraphics[width=0.85\textwidth]{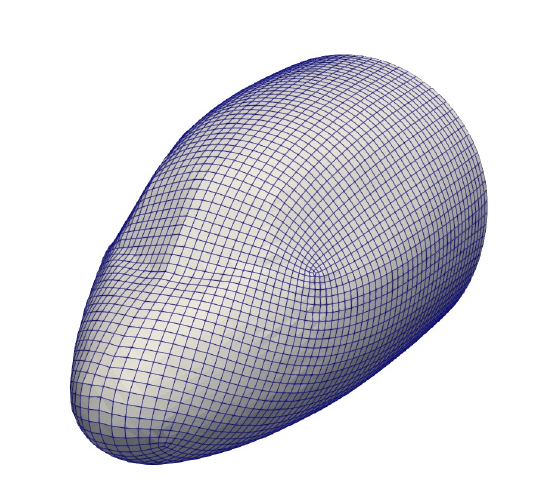}
            \caption{Quad mesh generation}
    \end{subfigure}
    \hfill
    \begin{subfigure}[b]{0.45\textwidth}
            \centering
            \includegraphics[width=0.75\textwidth]{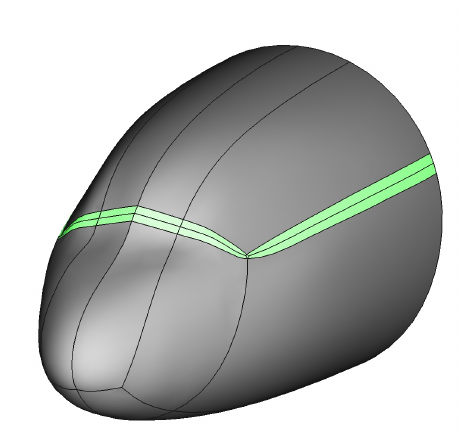}
            \caption{Patch extraction}
    \end{subfigure}
    \begin{subfigure}[b]{0.45\textwidth}
            \centering
            \includegraphics[width=0.7\textwidth]{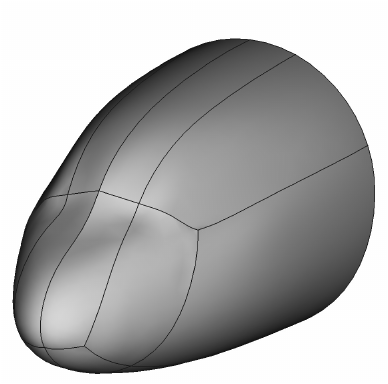}
            \caption{Patch simplification and fitting}
    \end{subfigure}
    \caption{Reconstruction of a plane head as a watertight muti-patch NURBS surface.}\label{fig14}
\end{figure}

The next model is a plane head, which is composed of 66 trimmed NURBS patches; see Fig.~\ref{fig14}(a). Both the topology and geometry are correct, so there is no need to perform mesh repair. A highly dense triangle mesh is generated with more than 16 million triangular faces, aiming to demonstrate the capability of our proposed pipeline in dealing with large-scale input (in terms of triangle meshes). The proposed pipeline is applied to this model. It finishes the entire reconstruction procedure in about 25 mins on a normal PC, demonstrating the efficiency of the proposed pipeline. 

\begin{figure}[h]
    \centering
    \begin{subfigure}[b]{0.45\textwidth}
           \centering
           \includegraphics[width=0.9\textwidth]{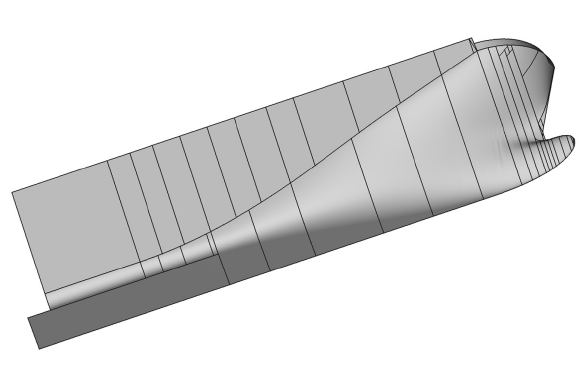}
            \caption{Trimmed CAD model}
    \end{subfigure}
    \hfill
    \begin{subfigure}[b]{0.45\textwidth}
            \centering
            \includegraphics[width=1.05\textwidth]{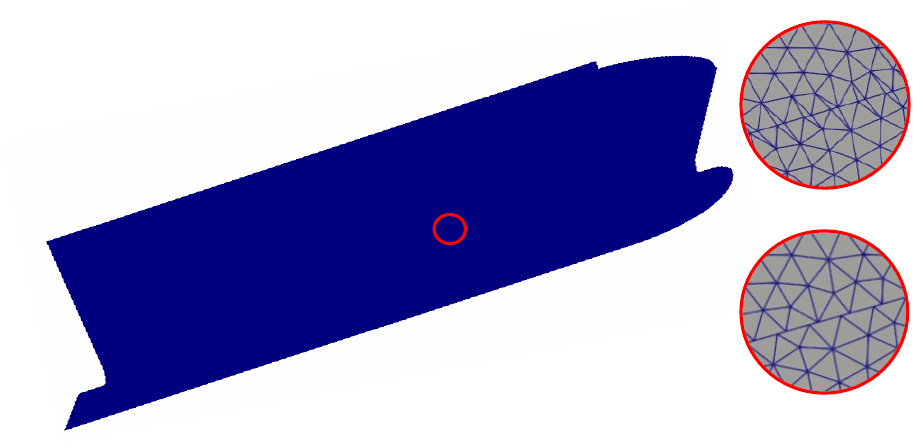}
            \caption{Triangulation and mesh repair}
    \end{subfigure}
    \begin{subfigure}[b]{0.45\textwidth}
           \centering
           \includegraphics[width=0.95\textwidth]{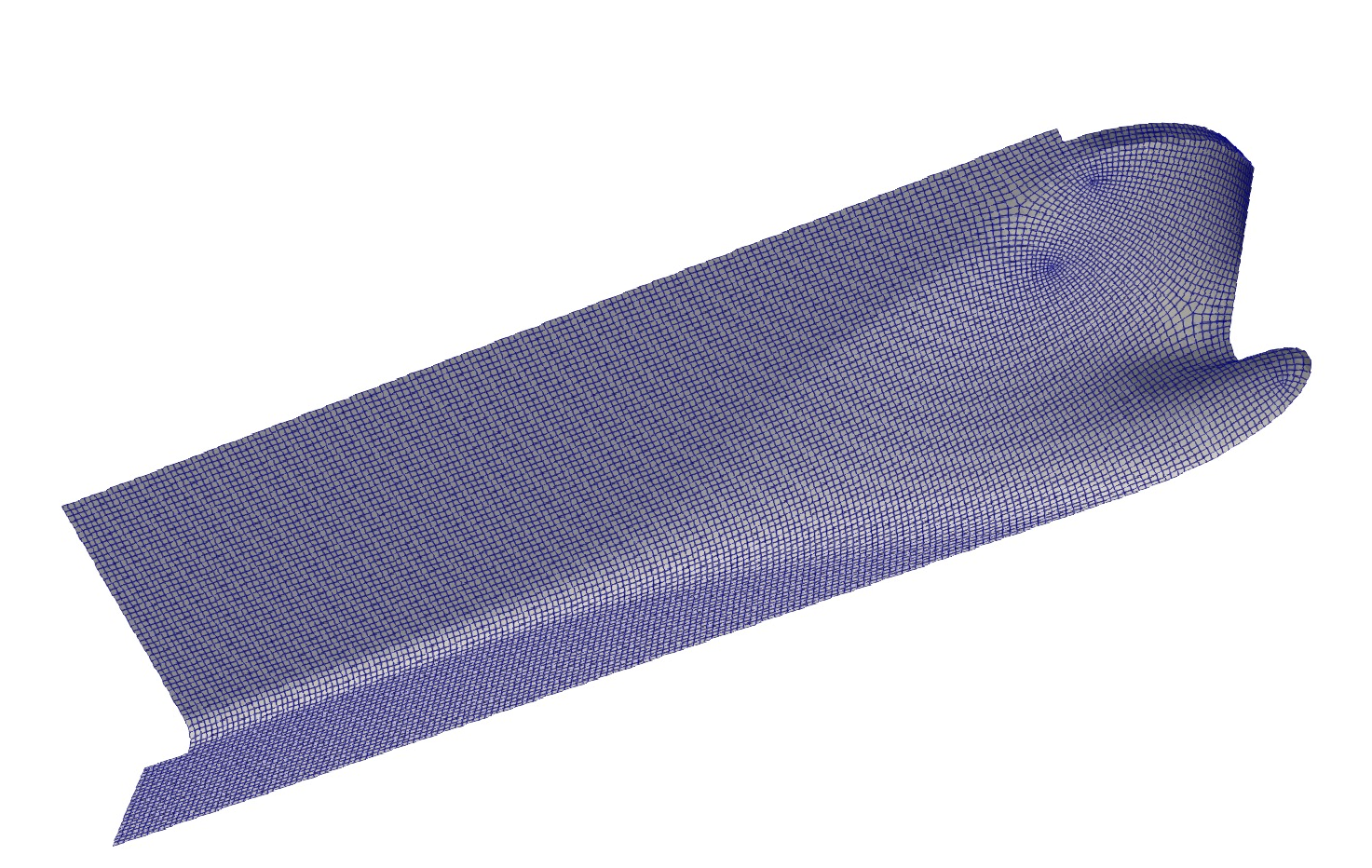}
            \caption{Quad mesh generation}
    \end{subfigure}
    \hfill
    \begin{subfigure}[b]{0.45\textwidth}
            \centering
            \includegraphics[width=0.87\textwidth]{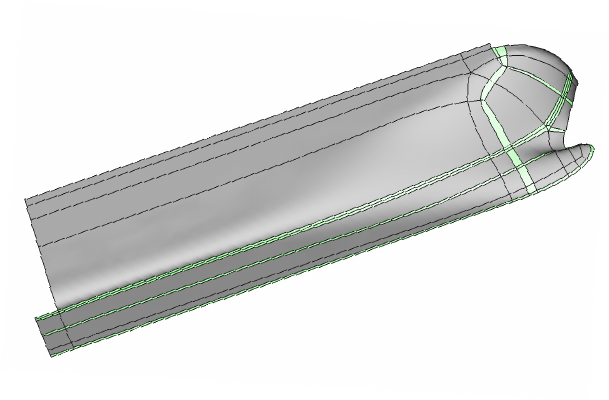}
            \caption{Patch extraction}
    \end{subfigure}
    \begin{subfigure}[b]{0.45\textwidth}
            \centering
            \includegraphics[width=0.85\textwidth]{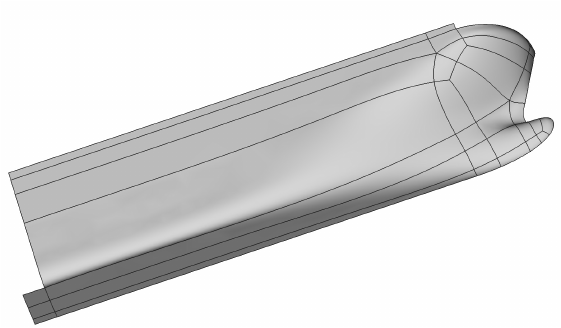}
            \caption{Patch simplification and fitting}
    \end{subfigure}
    \caption{Reconstruction of a ship hull as a watertight muti-patch NURBS surface.}\label{fig15}
\end{figure}

The last model is a ship hull, which has sharp corners on the boundary, internal sharp features, and drastically varying curvature. The topology of the input CAD model is not correct and thus it has invisible gaps. To demonstrate the proposed method capable of handling large-scale input and complex geometric features at the same time, a dense triangle mesh is generated with more than 11 million faces. The whole reconstruction procedure can be done in about 20 mins, with all the geometric features well preserved.

\begin{figure}[h]
    \centering
    \begin{subfigure}[b]{0.495\textwidth}
           \centering
           \includegraphics[width=1.1\textwidth]{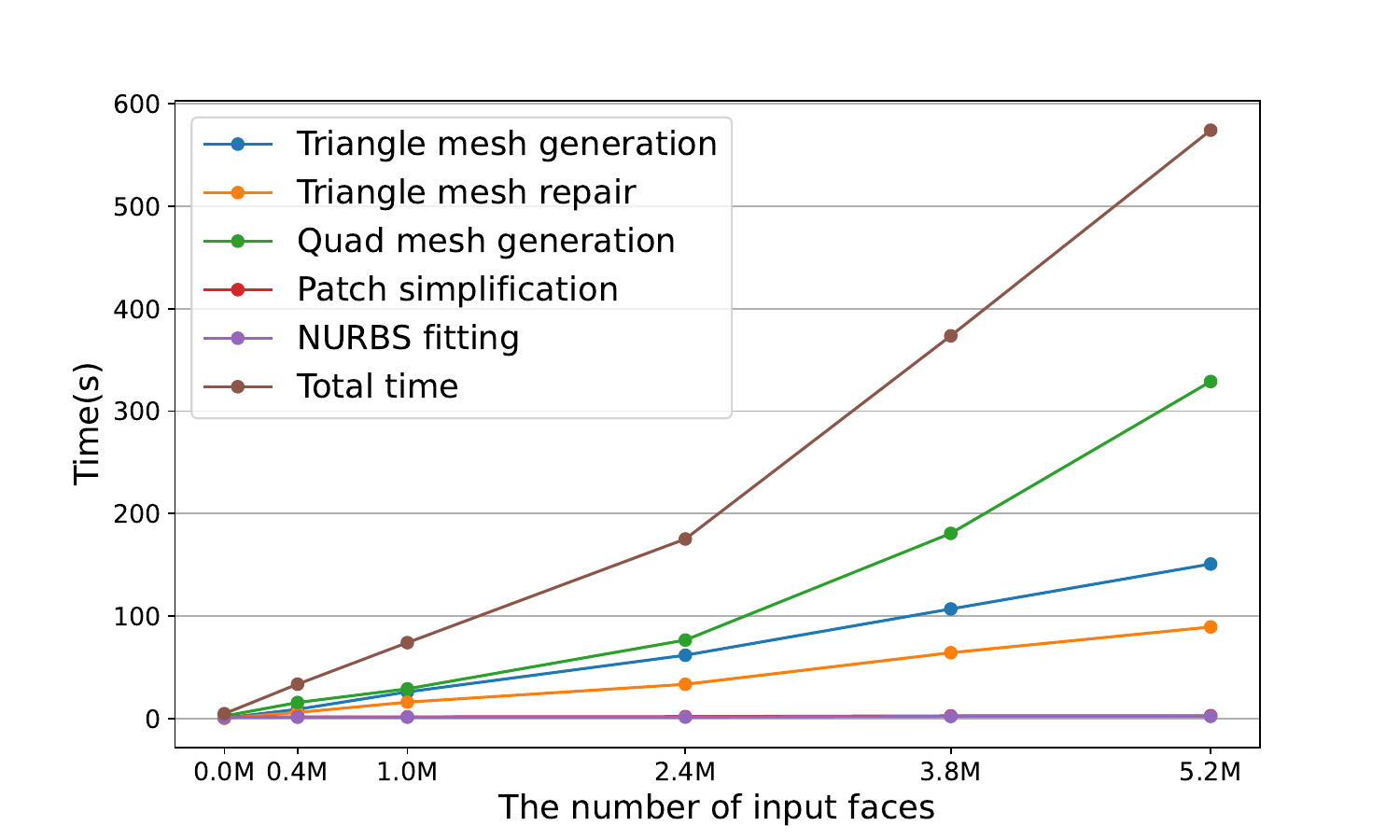}
            \caption{A plane plate model}
    \end{subfigure}
    \begin{subfigure}[b]{0.495\textwidth}
            \centering
        \includegraphics[width=1.1\textwidth]{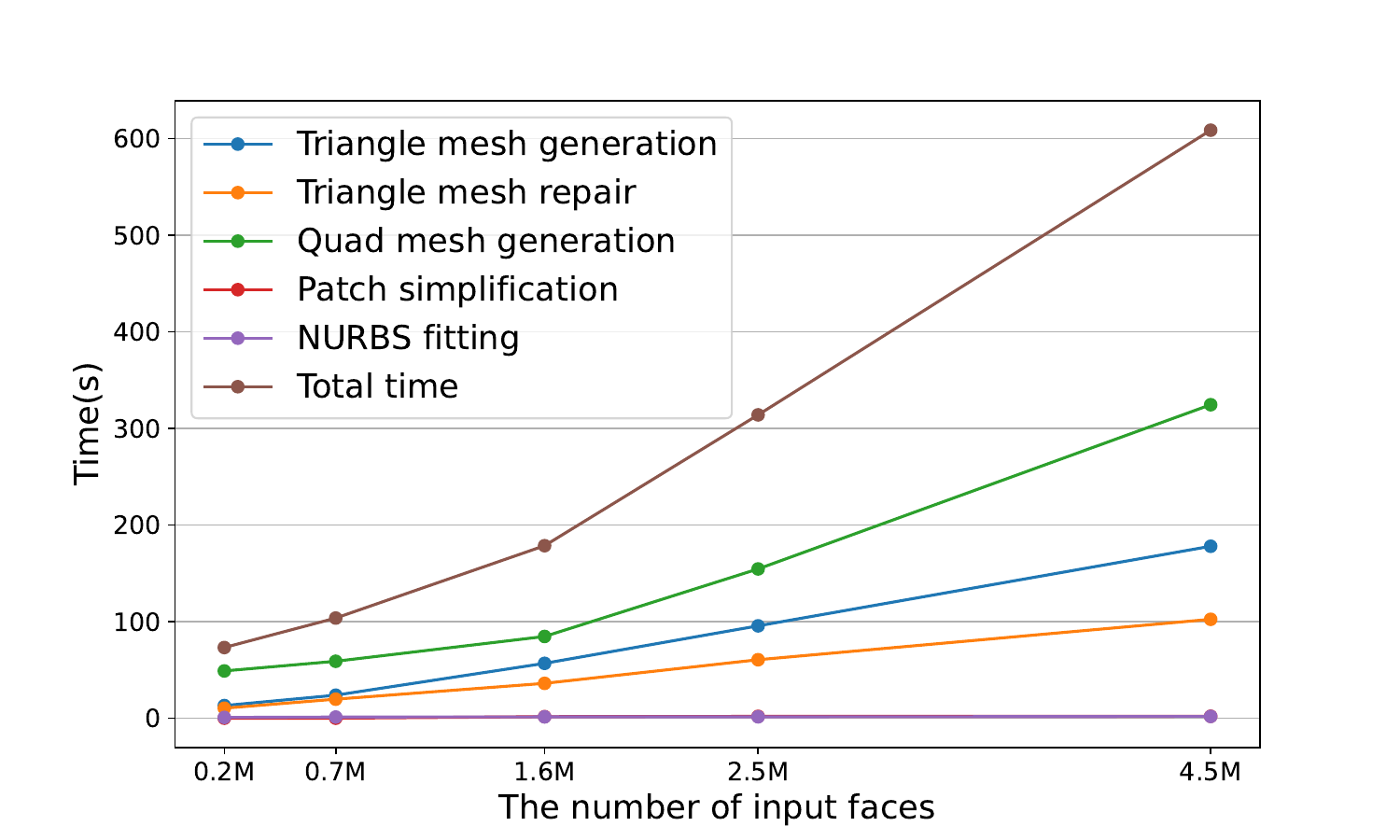}
            \caption{A ship head}
    \end{subfigure}
    \begin{subfigure}[b]{0.495\textwidth}
           \centering
           \includegraphics[width=1.1\textwidth]{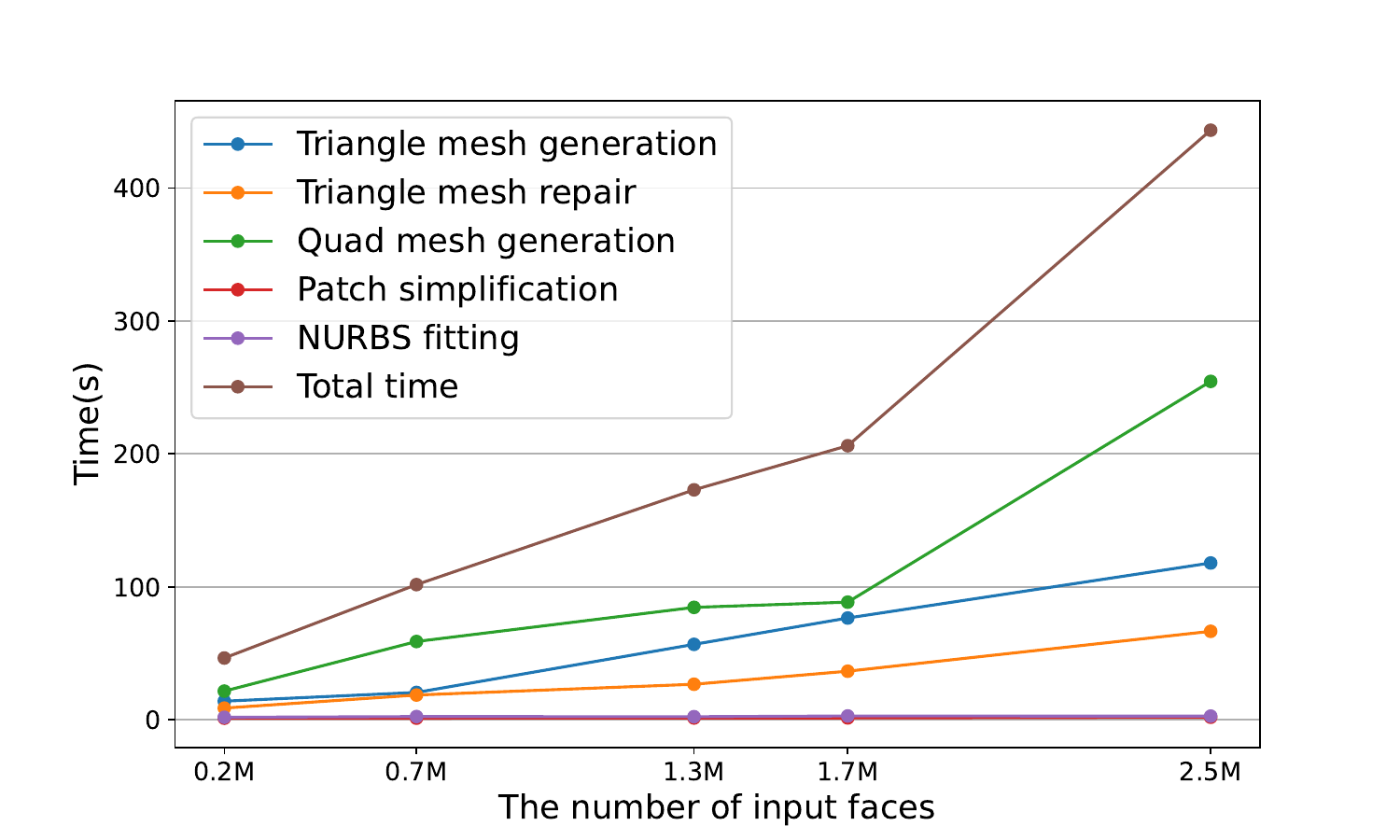}
            \caption{A car body}
    \end{subfigure}
    \begin{subfigure}[b]{0.495\textwidth}
            \centering
            \includegraphics[width=1.1\textwidth]{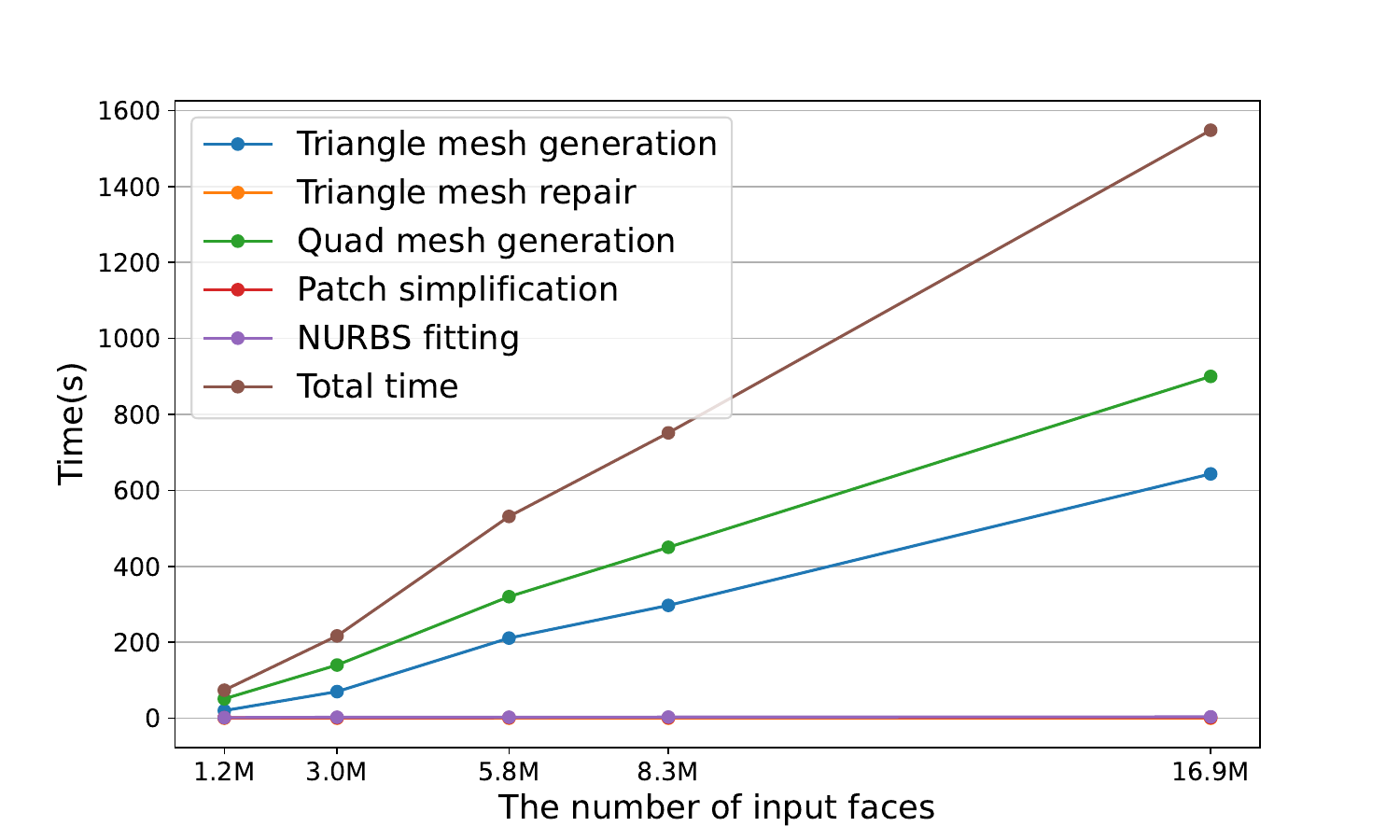}
            \caption{A plane head}
    \end{subfigure}
    \begin{subfigure}[b]{0.5\textwidth}
            \centering
            \includegraphics[width=1.1\textwidth]{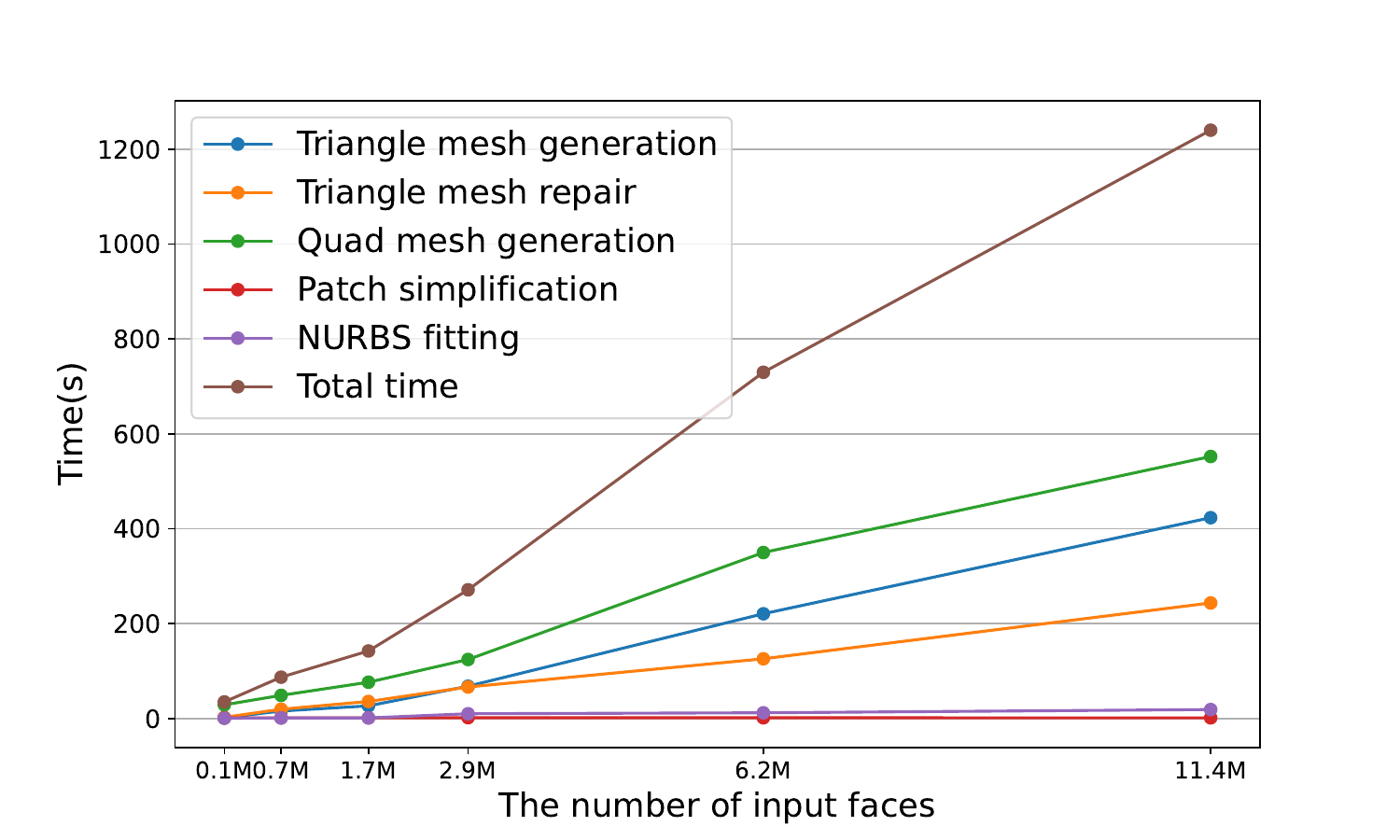}
            \caption{A ship hull}
    \end{subfigure}
    \caption{Computational time with respect to the size of input for each tested model.}\label{scalable}
\end{figure}

Last but not least, to show the scalability of the proposed method in more detail, for all the models we summarize the computational time with respect to the mesh size (in terms of triangular faces) for each key step and for the overall process; see Fig.~\ref{scalable}. We observe that the computational time for steps that cost the most (i.e., triangulation, mesh repair, and quad meshing) varies almost linearly with respect to the size of input (namely the number of triangular faces). As a result, the overall computational time (brown lines) varies (almost) linearly with respect to the size of input, demonstrating the scalability of the proposed method.

\subsection{Isogeometric analysis}
With a watertight representation, we can use it to perform isogeometric analysis (IGA) through B\'ezier extraction~\cite{borden2011isogeometric}. B\'ezier extraction provides a handy way to integrate spline representations with existing codes based on the finite element method (FEM) through a common exchange format, where each spline function, when restricted to the element level, is represented as a linear combination of Bernstein polynomials that are fairly easy to evaluate. The reconstructed surface models naturally correspond to shell structures, so we perform IGA on such shells using our in-house code, which is based on the linear Kirchhoff-Love shell. 

 \begin{figure}[h]
  \subcaptionbox{Boundary conditions}[0.45\textwidth]{\includegraphics[width=0.8\linewidth]{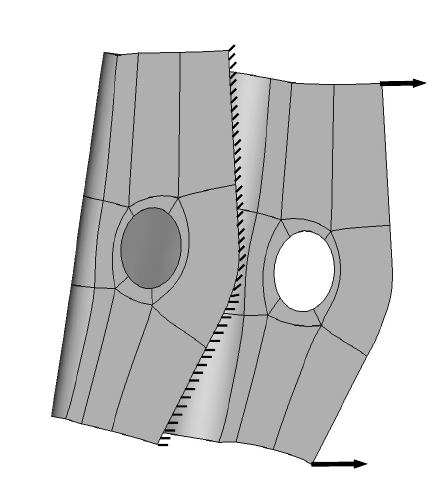}}
  \hfill
  \subcaptionbox{Deformation results}[0.46\textwidth]{\includegraphics[width=0.9\linewidth]{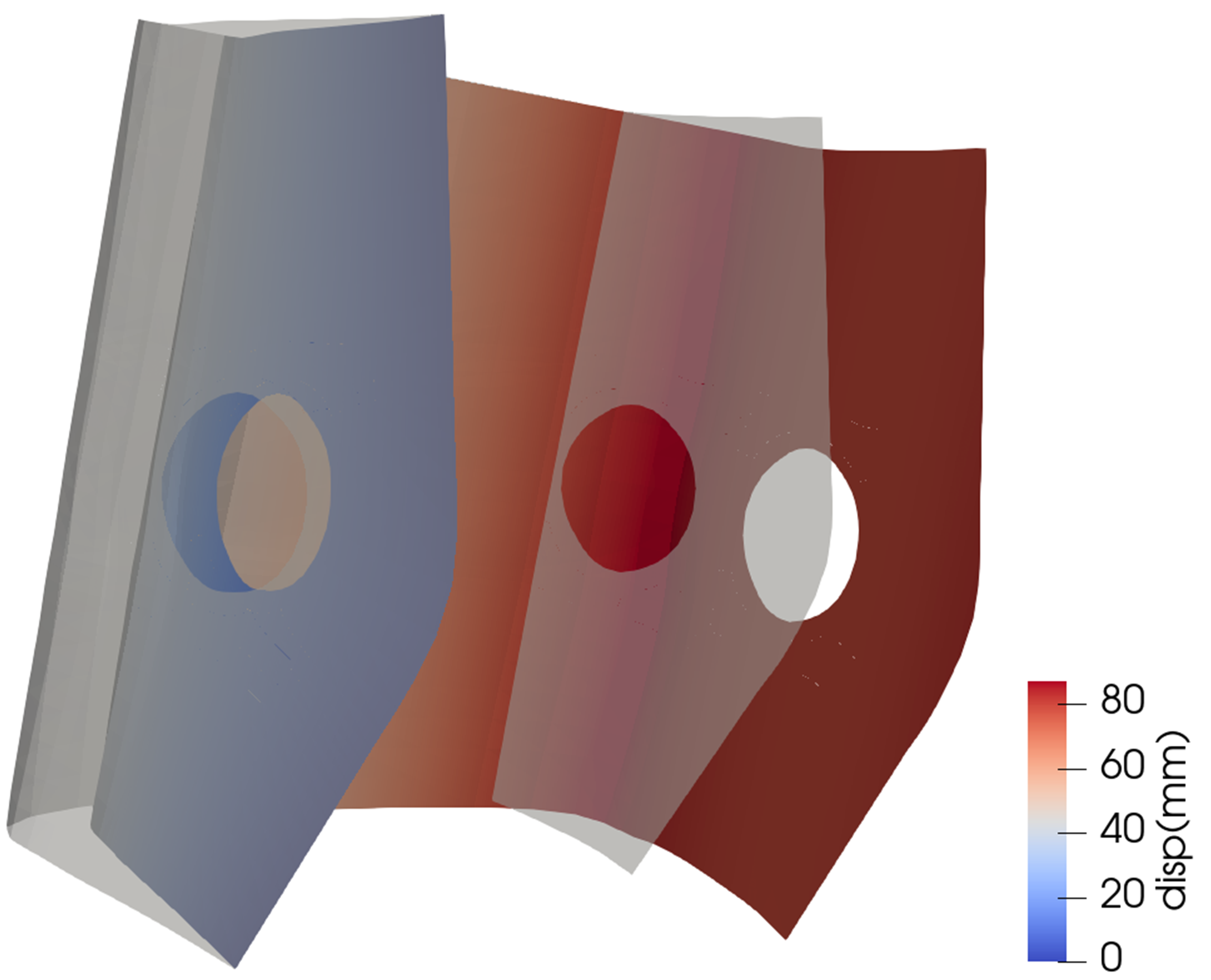}}
  \caption{Isogeometric analysis on the plane plate model. (a) Boundary conditions, and (b) the deformation result, where the light gray shade represents the undeformed shape.}
  \label{simulation2}
\end{figure}

% \begin{figure*}[t]
%     \centering
%     \begin{subfigure}[b]{0.45\textwidth}
%            \centering
%            \includegraphics[width=0.96\textwidth]{ship_simulation_a.pdf}
%            \caption{Boundary conditions}
%     \end{subfigure}
%     \hspace{1cm}
%     \begin{subfigure}[b]{0.46\textwidth}
%             \centering
%             \includegraphics[width=1\textwidth]{ship_simulation_b.png}
%             \caption{Deformation results}
%     \end{subfigure}
%     \caption{Isogeometric analysis on the ship hull model. (a) Boundary conditions, and (b) the deformation result, where the light gray shade represents the undeformed shape.}\label{simulation1}
% \end{figure*}

 \begin{figure}[h]
  \subcaptionbox{Boundary conditions}[0.45\textwidth]{\includegraphics[width=0.96\linewidth]{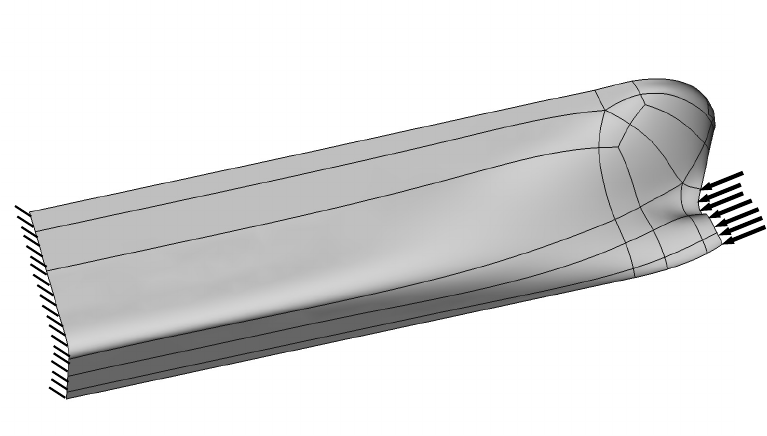}}
  \hfill
  \subcaptionbox{Deformation results}[0.46\textwidth]{\includegraphics[width=1\linewidth]{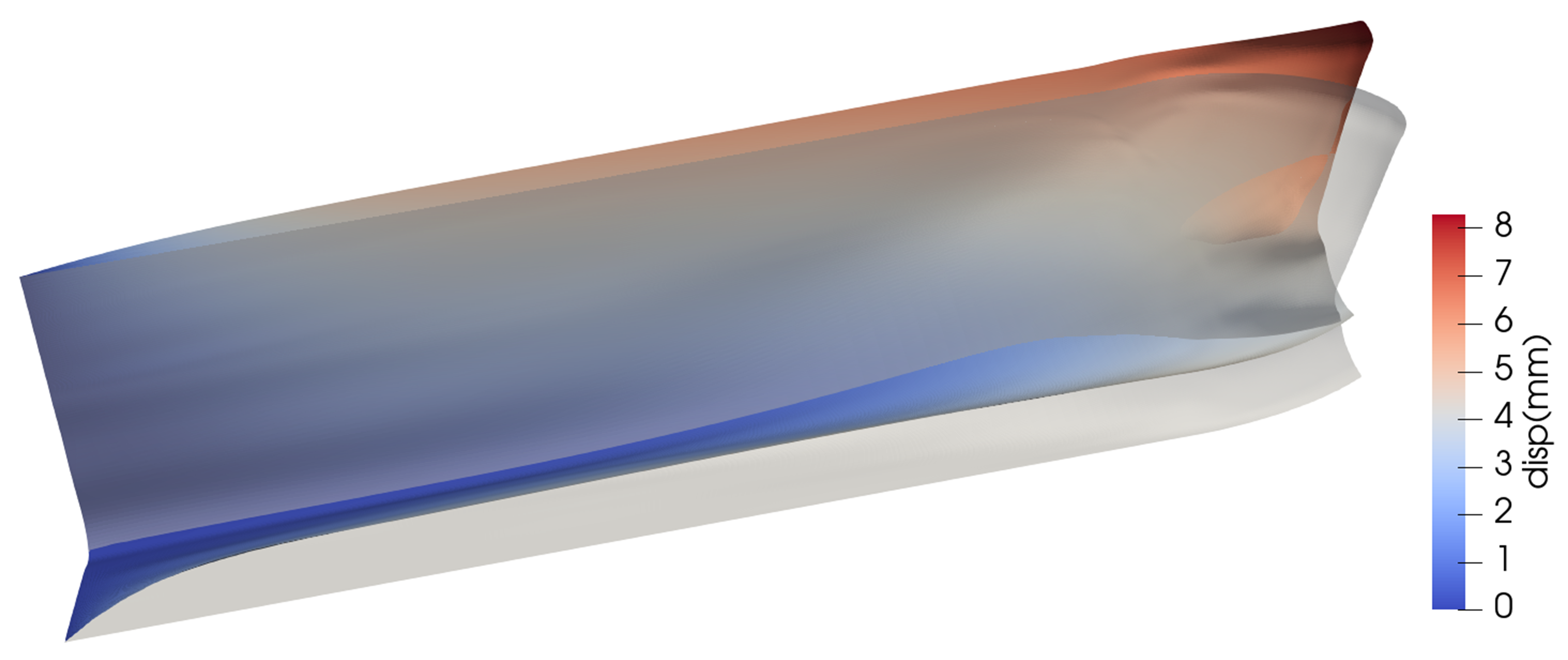}}
  \caption{Isogeometric analysis on the ship hull model. (a) Boundary conditions, and (b) the deformation result, where the light gray shade represents the undeformed shape.}
  \label{simulation1}
\end{figure}

The Kirchhoff-Love shell requires basis functions that are globally \(C^1\)-continuous, whereas the reconstructed models (based on multi-patch NURBS) are only \(C^0\)-continuous across patch interfaces, so they cannot be applied directly. On the other hand, the multi-patch NURBS has a uniform parameterization by construction. We can take advantage of its control mesh, which is a coarse quad mesh with EPs, and construct unstructured splines that are at least \(C^1\)-continuous everywhere~\cite{ref:wei22, ref:casquero20}.

We perform IGA on two models. First, we study the plane plate model. A linear material model is used, with the Poisson's ratio \( \nu = 1/3 \) and Young's modulus \( E = 2.06 \times 10^{11} \) Pa. The boundary conditions are shown in Fig.~\ref{simulation2}(a), where the concentrated force has a magnitude of \( F = 2 \times 10^8 \) N. The deformation result is shown in Fig.~\ref{simulation2}(b), demonstrating the analysis-suitability of the watertight spline representation.

Second, we study the ship hull model. The two material parameters are the same as in the previous example. Boundary conditions are given in Fig.~\ref{simulation1}(a), where the left boundary is clamped and a distributed load is imposed at the right bottom with a magnitude \( F = 1 \times 10^7 \) N. The deformation result is shown in Fig.~\ref{simulation1}(b).

% \begin{figure*}[h]
%     \centering
%     \begin{subfigure}[b]{0.45\textwidth}
%            \centering
%            \includegraphics[width=0.8\textwidth]{plane_plate_simulation_a.pdf}
%              \caption{Boundary conditions}
%     \end{subfigure}
%     \hspace{1cm}
%     \begin{subfigure}[b]{0.46\textwidth}
%             \centering
%             \includegraphics[width=0.9\textwidth]{plane_plate_simulation_a.png}
%             \caption{Deformation results}
%     \end{subfigure}
%     \caption{Isogeometric analysis on the plane plate model. (a) Boundary conditions, and (b) the deformation result, where the light gray shade represents the undeformed shape.}\label{simulation2}
% \end{figure*}

\section{Conclusion and future work}\label{con}
In this paper, we present a modularized, semi-automatic, and scalable pipeline to reparameterize trimmed CAD models entirely as watertight spline representations. The pipeline consists of several key steps, including triangulation, mesh repair, quad meshing, patch extraction and simplification, and spline fitting. Among them, triangulation and quad meshing often dominate the computational cost. While there is little we can do about triangulation in this work (as we use Gmsh and OpenCASCADE for this purpose), QuadriFlow-based quad meshing relies on local operators to achieve high-quality and field-aligned parameterizations, thus enabling scalable algorithms that can handle large-scale input triangle meshes. On top of QuadriFlow, we impose boundary constraints to support open surfaces for real-world engineering applications based on shells. While QuadriFlow keeps the number of singularities small, their placement is not guaranteed to be optimal, leading to redundant tiny and slender patches. Therefore, we propose a dedicated patch simplification method to eliminate such patches and thus achieve a much cleaner quad layout. Finally, the proposed pipeline is tested upon a variety of complex shell structures, demonstrating its capability to efficiently deal with large-scale inputs and complex geometries with various features.

In the future, we will improve the pipeline primarily in terms of accuracy and adaptivity. First, right now the relative fitting error of the reconstructed model is limited to about \(10^{-3}\). This error limit is determined by triangulation and quad meshing in an entangled manner. It can pose quite a practical issue when the size of the model is big, leading to a substantial deviation of the reconstructed model from the original model. Second, right now the quad mesh is always uniform. While it is appealing in many applications, engineering problems with local features are almost everywhere. Therefore, adaptive meshing based on QuadriFlow will be another promising research direction to pursue.

\bmhead{Acknowledgements}
Z. Wei and X. Wei are partially supported by National Natural Science Foundation of China (No. 12202269).

\begin{appendices}
\section{Triangle mesh repair}\label{app:fixing}

\begin{figure}[h]
    \centering
    \begin{subfigure}[b]{0.32\textwidth}
           \centering
           \includegraphics[width=1\textwidth]{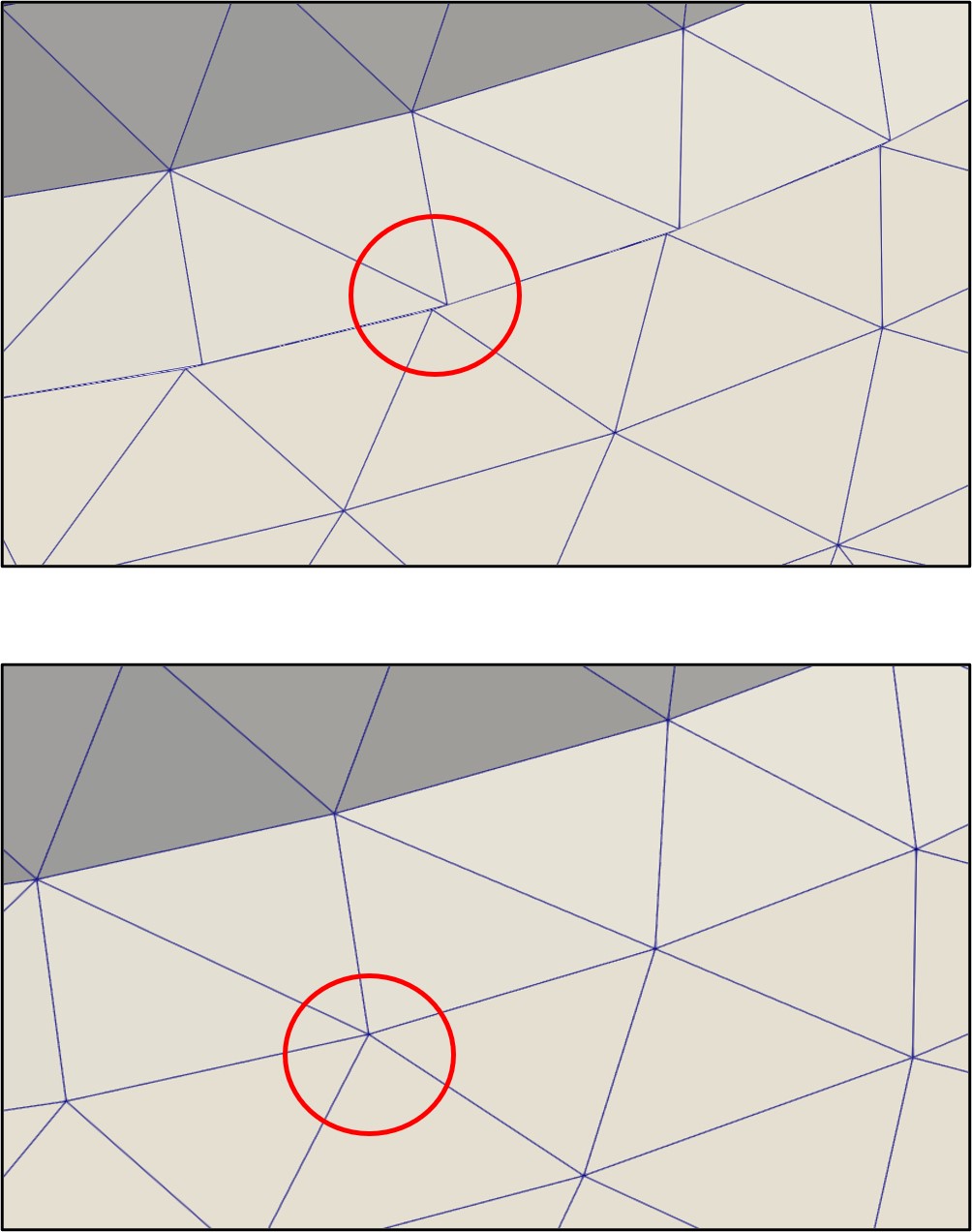}
           \caption{Vertec-vertex mismatch}
    \end{subfigure}
    \begin{subfigure}[b]{0.32\textwidth}
            \centering
            \includegraphics[width=1\textwidth]{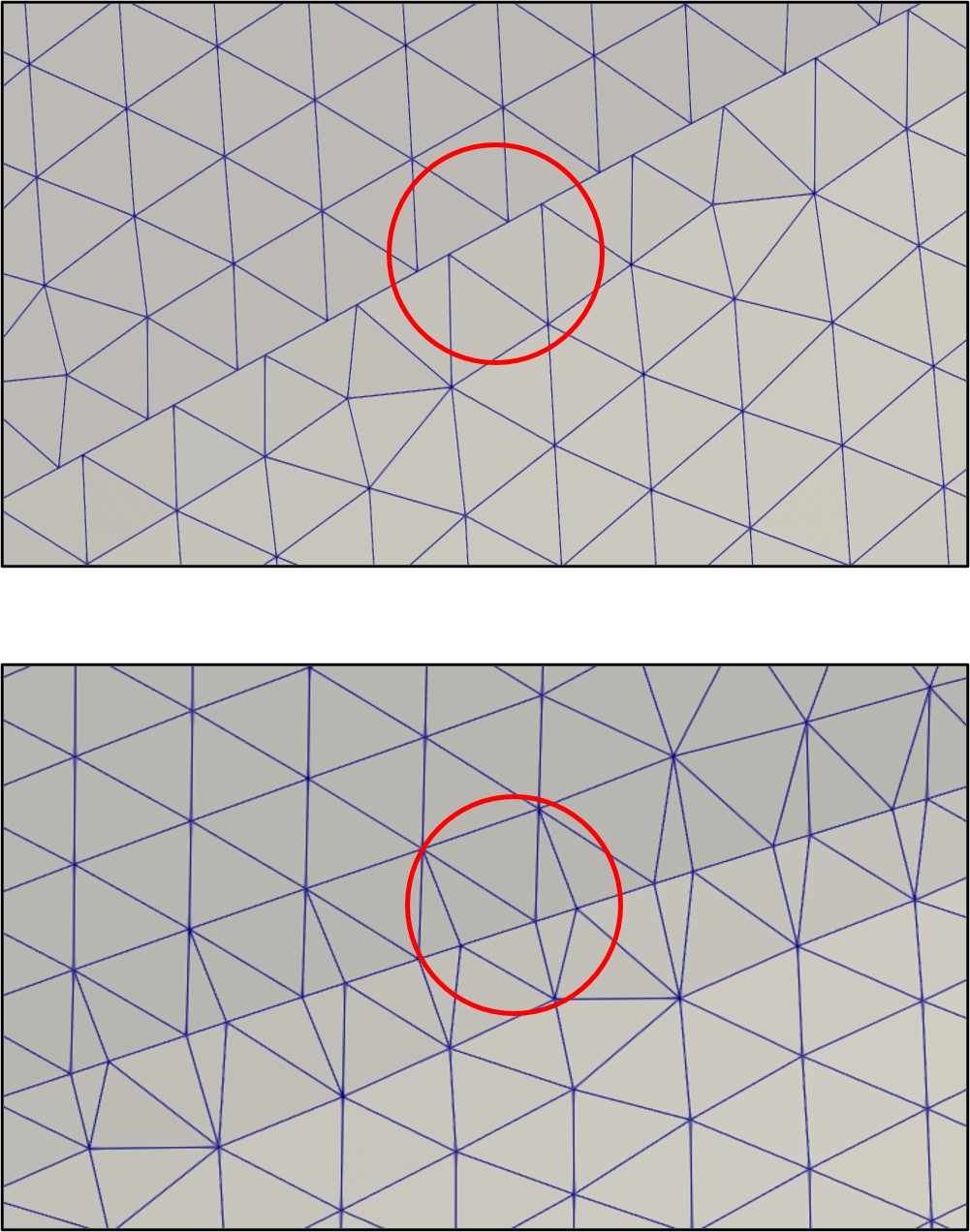}
            \caption{Vertec-edge mismatch}
    \end{subfigure}
     \begin{subfigure}[b]{0.32\textwidth}
            \centering
            \includegraphics[width=1\textwidth]{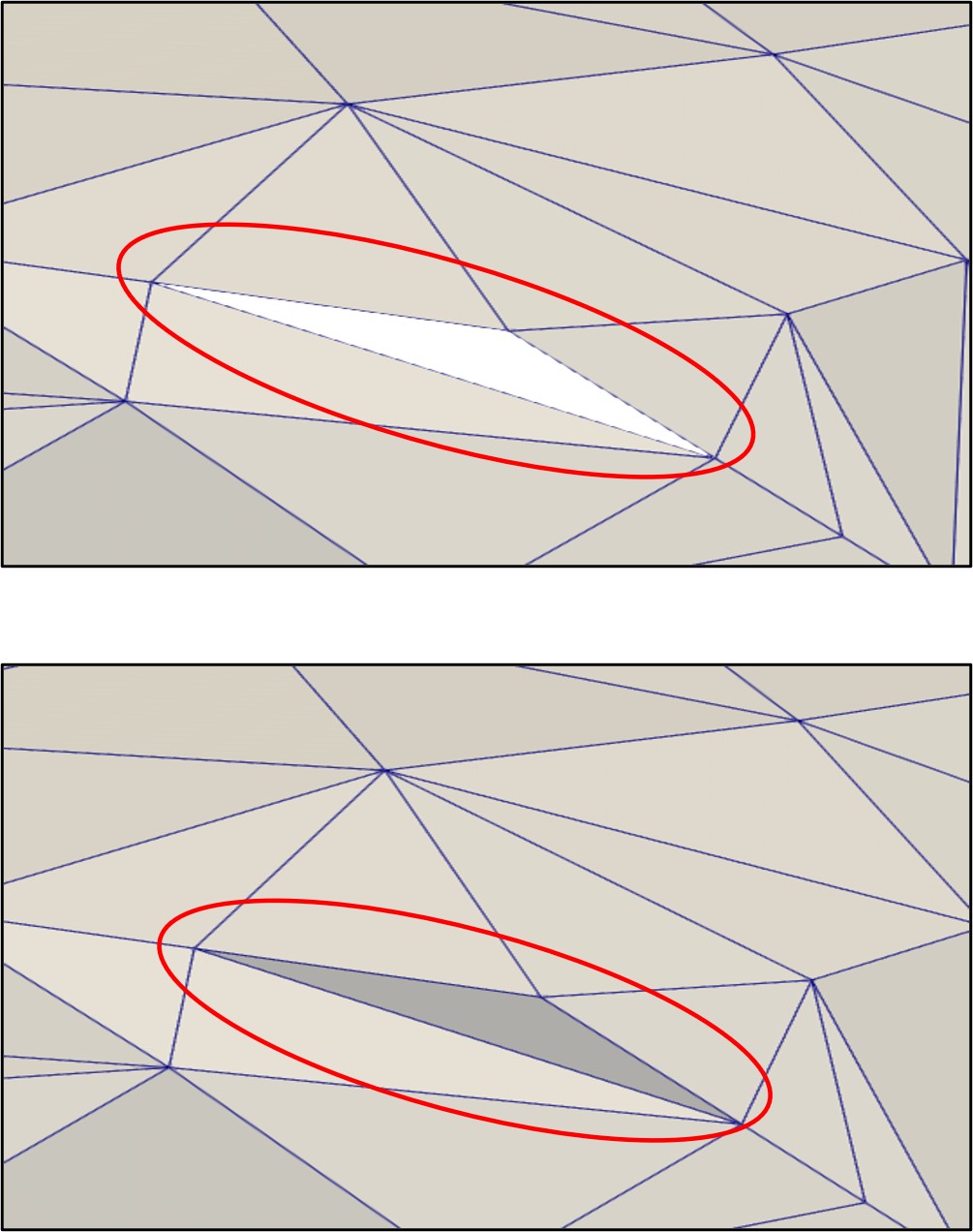}
            \caption{Large gap repair}
    \end{subfigure}
    \caption{Three common cases that need mesh repair. }\label{trifix}
\end{figure}

Once an initial triangle mesh is obtained by triangulating the CAD model, the proposed pipeline proceeds with a mesh-repair step to address the trimming-related issues. Trimming can lead to various problems, such as self-intersections, gaps, and overlaps. In this work, we focus on dealing with undesired gaps, a typical scenario in practical applications. As shown in Fig.~\ref{trifix}, we perform dedicated treatments for three involved cases. All the cases occur only on the patch boundaries.
\begin{enumerate}[label=$\bullet$, topsep = 7 pt]
\item Vertex-vertex mismatch. When two vertices are identified to share the same position within a user-defined tolerance (e.g., less than one-fifth the length of the shortest edge connected to the vertex), it means that there are duplicated vertices in the mesh; see Fig.~\ref{trifix}(a). In this case, we simply remove one of them and update the corresponding index.
\item Vertex-edge mismatch. When the distance between a vertex and an edge is smaller than a given tolerance, the vertex is said to lie on the edge and we have the case of vertex-edge mismatch; see Fig.~\ref{trifix}(b). In this case, we split the edge into two edges that are connected by the vertex. Subsequently, the face sharing the edge is also split into two faces.
\item Large gap. The size of a gap may exceed the given tolerance; see Fig.~\ref{trifix}(c). When this is the case,  we fill the gap with a triangle mesh whose vertices coincide exactly with those of the gap.
\end{enumerate}

There are many other possibilities related to gaps and overlaps, which, however, go beyond the scope of this work. Interested readers may refer to~\cite{ju2009fixing} for a comprehensive discussion on the topic.

\section{NURBS fitting}\label{app:fitting}
Once a well-structured quad mesh is obtained, the final step of the proposed pipeline is to fit a spline surface to the quad mesh. For this we take full advantage of the multi-patch structure that is already in place after patch simplification, where each patch corresponds to a regular quad mesh. Usually, the quad mesh itself is dense and not suited to serve as the control mesh for a spline model, so spline fitting is needed.

Spline fitting is done in a patch-by-patch manner and it involves two steps. First, every patch of quad mesh is parameterized individually using the Floater's algorithm~\cite{ref:floater97, ref:floater03}, as every patch has a rectangular parametric domain. As a result, every point in the quad mesh has a pair of parametric coordinates. Second, B-spline fitting is performed through the least-squares approximation~\cite{ref:nurbsbook}. By default, we take $10^{-3}$ as the tolerance for the relative fitting error, i.e., the maximum derivation normalized by the diagonal length of the bounding box. The process of surface fitting is iterated until the error is smaller than the given tolerance. However, sometimes the error may not be reduced further because of the limited resolution of the quad mesh. When this is the case, a finer quad mesh can be quickly generated and spline fitting will be applied to this finer mesh.

To maintain the conformality between adjacent spline patches, we adopt the same degree for both parametric directions and start with a conformal multi-patch control mesh, e.g., a $2\times 2$ control mesh for every patch. During the iteration of surface fitting, we perform global refinement in every patch at the same time when more control points are needed to bring the fitting error down. This guarantees that the number and the positions of control points on every patch interface always coincide.
\end{appendices}


\bibliography{reference}  


\end{document}